\documentclass[%
 reprint,
 amsmath,amssymb,
 aps,
]{revtex4-1}

\usepackage{graphicx}
\usepackage{dcolumn}
\usepackage{bm}

\usepackage{tikz}
\usepackage{pgfplots}

\begin{document}

\title{Elementary Cellular Automata along with delay sensitivity can model communal riot dynamics}

\author{Souvik Roy}
\email{svkr89@gmail.com}
\altaffiliation[]{Department of Information Technology, Indian Institute of Engineering Science and Technology, Shibpur, Howrah, West Bengal, India 711103.}
\author{Abhik Mukherjee}
\email{abhikm.kol@gmail.com}
\altaffiliation[]{Computer Science and Technology, Indian Institute of Engineering Science and Technology, Shibpur, Howrah, West Bengal, India 711103.}
\author{Sukanta Das}
\email{sukanta@it.iiests.ac.in}
\altaffiliation[]{Department of Information Technology, Indian Institute of Engineering Science and Technology, Shibpur, Howrah, West Bengal, India 711103.}

\date{\today}


\begin{abstract}
This work explores the potential of elementary cellular automata to model the dynamics of riot. Here, to model such dynamics, we introduce probabilistic loss of information and delay perturbation in the updating scheme of automata to capture sociological parameters - presence of anti-riot population and organizational presence of communal forces in the rioting society respectively. Moreover, delay has also been incorporated in the model to capture the non-local interaction of neighbours. Finally, the model is verified by a recent event of riot that occurred in Baduria of West Bengal, India.
\end{abstract}

\pacs{Valid PACS appear here}
\maketitle
\section{Introduction}
Riots and their dynamics have been a popular topic to sociologists and historians \cite{midlarsky1978,Rodger26,Bohstedt21,charlesworth1994,Daniel4v,Myers57454,
Das7477,das1993communal,article1}. In a parallel journey, computer scientists and mathematicians have found their interest in the study of riots, with a target to mathematically model their dynamics \cite{9989878,Laurent211,Davies2013211,Berestycki11,Brian9,Daniel10,Braha8,
BAUDAINS2013211,2778111,10.2307/2094159}. The most popular approach of developing such models is to adopt epidemiological framework \cite{9989878,Laurent211,Davies2013211,Berestycki11,BAUDAINS2013211}. For example, $1960$'s Los Angeles ($1965$) - Detroit ($1967$) - Washington D.C. ($1968$) riots \cite{9989878}, $2005$ French riot \cite{Laurent211}, $2011$ London riots \cite{Davies2013211} etc. have been modelled using this approach. Recently, non-local interactions along with neighbourhood dependency of elements of the system have been introduced in the models of riots \cite{Berestycki11,Braha8,Laurent211,Sandra11}. It is argued that due to the globalization and the advent of communication technology, long range, that is, non-local communications among elements are necessary to better model the dynamics of riots.

In this scenario, we undertake this research to show that the elementary cellular automata (ECAs) which rely only on local neighbourhood dependency can efficiently model the dynamics of riots, if the neighbourhood dependency is {\em delay sensitive}. In particular, to model riot dynamics by ECAs, we introduce `probabilistic loss of information perturbation' and `delay perturbation' in the updating scheme of the automata. We observe that due to this updating scheme, the ECAs show a new kind of dynamical behaviour, which suggests us that some ECAs can be better model of riot. Finally, to validate our claim, we take into consideration a recent riot that happened in Baduria of West Bengal, India. Since media reports do not always reflect the ground realities of riots, we organized an extensive field study in Baduria to get insight about rioting dynamics. 

Here, in the proposed ECA based model, probabilistic loss of information perturbation rate is related to sociological factor such as the presence of {\em anti-riot} population in the rioting society. Similarly, the presence of communal elements in society, which plays a role to regenerate rioting spontaneity in the rioting society, indicates the physical implication of delay in the system. However, the inherent property of CA is local interaction which contradicts the recent trends of considering non-locality in the age of globalization \cite{Berestycki11,Braha8,Laurent211,Sandra11}. The delay passively induces a non-locality in the environment. To illustrate, the updated state information of a cell at time $t$ reaches to its neighbouring cell at $t+n$ time step where $n$ depicts delay for the cell and its neighbouring cell. This implies, non-local information from distance $n$ reaches to the corresponding neighbouring cell. The presence of communal organization in society, which physically indicates delay in the system, induces this non-locality to regenerate the rioting spontaneity.

\section{Delay sensitive cellular automata}

Here, we work with simple one-dimensional three-neighbouring two-state cellular automata, which are commonly known as elementary cellular automata (ECA) \cite{wolfram86}. The next state of each CA cell is determined as $\mathcal{S}_i^{t+1}$ = $f$($\mathcal{S}_{i-1}^t$,$\mathcal{S}_i^t$,$\mathcal{S}_{i+1}^t$) where $f$ is the next state function, $\mathcal{S}_{i-1}^t$,$\mathcal{S}_i^t$,$\mathcal{S}_{i+1}^t$ are the present states of left, self and right neighbour of the $i$-th CA cell at time $t$, respectively. The local transition function $f$: \{$0,1$\}$^3 \rightarrow$ \{$0,1$\} can be expressed as eight arguments of $f$. The decimal counterpart of eight next state is referred as `rule'. Each rule is associated with a `decimal code' $w$, where $w$ =$f$($0,0,0$).$2^0$ + $f$($0,0,1$).$2^1$ + $\cdots$ + $f$($1,1,1$).$2^7$, for the naming purpose. There are $2^8$($256$) ECA rules, out of which $88$ are minimal representative rules and the rest are their equivalent \cite{Li90thestructure}. Classically, all the cells of a CA are updated simultaneously. In the last decade, researchers have explored dynamics of CA under asynchronous updating schemes \cite{jcaFates14,ROY2019600,SethiRD16,SCHONFISCH1999123,PhysRevE5976,
PhysRevE910421,Bou2012,130519}.

Classically, in ECA, delay and probabilistic loss of information during information sharing among the neighbouring cells is not considered. In traditional cellular automata, if a cell updates its state at time $t$, then that state information is available to neighbouring cell at time $t+1$. To define the delay involved in sharing of information for two neighbouring cells $i$ and $j$ ($i \neq j$), we introduce a non-negative integer function D($i$,$j$). In the proposed system, D($i$,$j$) = D($j$,$i$) $\geqslant 1$ for any pair of neighbouring cells $i$ and $j$. To illustrate, D($i$,$j$) = $n$ in the system implies, if cell $i$ updates its state at time $t$, then the updated state information is available to cell $j$ at time $t+n$. In the proposed system, the delays are non-uniform in space; i.e. D($i$,$j$) may be different from D($i'$,$j'$), where $i$ and $j$; $i'$ and $j'$ are neighbouring cells, however, the delays are uniform in time. Practically, the delay perturbation parameter $d \in \mathbb{N}$ assigns the maximum possible delay in the proposed CA system. Every pair of neighbouring cells are randomly initialized with delay between $1$ to $d$ following a uniform distribution. For the loss of information, one can consider that the delay is $\infty$ (infinity). Here, $\iota$ ($0 \leq \iota \leq  1$) indicates the probabilistic loss of information perturbation rate.

Now, for introducing probabilistic loss of information and delay in the system, each cell has to maintain state information of neighbours to get a view of neighbour's state. In the proposed system, each cell has a {\em view} about the states of its neighbours which may change from time to time depending on the arrival of state information about neighbours. However, the cells act depending on the current state information about neighbours at that time. In this context, the state set is distinguished into two parts - the actualstate (self) of a cell and a vector of neighbour's viewstate. Now, the state set can be written as $\mathcal{S'}$ = $\mathcal{S} \times \mathcal{S}^2$. Therefore, for a cell $c$, configuration at time $t$ is distinguished into two parts - $a_c^t$ and $\textbf{v}_c^t$ where $a_c^t \in \mathcal{S}$ is the actualstate and $\textbf{v}_c^t \in \mathcal{S}^2$ is the vector of viewstate of left and right neighbours. Note that, the actualstate set $\mathcal{S}$ is sufficient to represent traditional CA. Here, in the proposed CA system, the local transition function is also sub-divided into two parts - in the first {\em state update} step, a cell changes its {\em actualstate} depending on the {\em actualstate} of the self and {\em viewstates} of neighbours; and, in the second {\em information sharing} step, the cell shares its updated {\em actualstate} to its neighbouring cells. Now, the local transition function can be written as $f'$ = $f_u \circ f_s$, where, $f_u$ is the {\em state update} function, and $f_s$ is the {\em information-sharing} function. Here, the operator `$\circ$' indicates that the functions are applied sequentially to represent the actual update.

\begin{figure}[!htbp] 
\centering 
\includegraphics[width=2.1in]{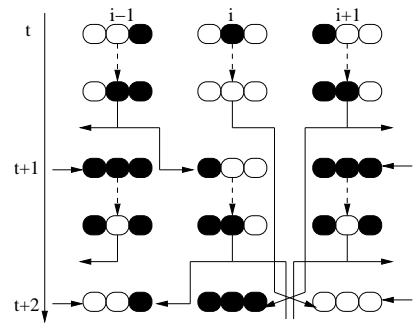} 
\caption{Example of delay and probabilistic loss of information perturbation updating scheme. The applied rule is ECA $50$.}
\label{ex1}
\end{figure}

To illustrate, Fig~\ref{ex1} depicts a simple $3$-cell ECA, where $D$($i$-$1$,$i$) = $D$($i$+$1$,$i$-$1$) = $1$ and $D$($i$,$i$+$1$) = $2$. In Fig.~\ref{ex1}, each cell has a view about the states of neighbours, i.e. left and right one for each cell. For every time step, the first step (dotted line) shows the state update function and the second step (straight line) shows the information sharing function. Here, the information about state change of cell $i$ (resp. cell $i$+$1$) at first time step reaches to cell $i$+$1$ (resp. cell $i$) at third time step due to delay perturbation. In Fig~\ref{ex1}, the information about state change of cell $i$ at first time step does not reach to cell $i$-$1$ at second time step due to probabilistic loss of information perturbation. 

To sum up, the proposed CA system depends on the following two parameters : (i) The delay perturbation parameter $d \in \mathbb{N}$ indicates the maximum delay limit of the system; (ii) The probabilistic loss of information perturbation rate $\iota$ ($0 \leq \iota \leq 1$) indicates the probabilistic loss of information during information sharing. 

\section{Modeling of riots}
\subsection{Dynamic behaviour}

\begin{figure*}
\begin{tabular}{cccc}
ECA 14 & ECA 60 & ECA 30 & ECA 18 \\[6pt]
  \includegraphics[width=35mm]{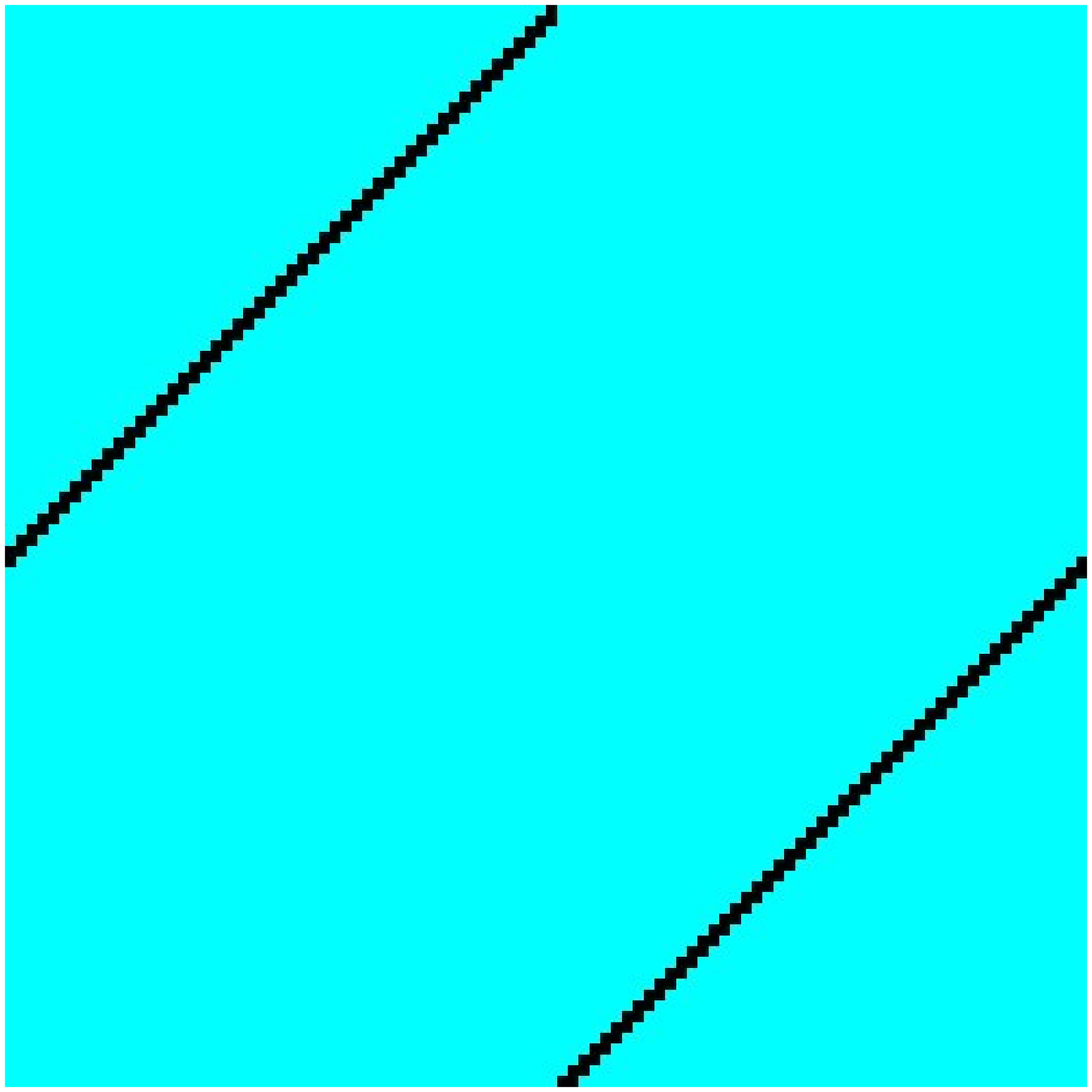} & \includegraphics[width=35mm]{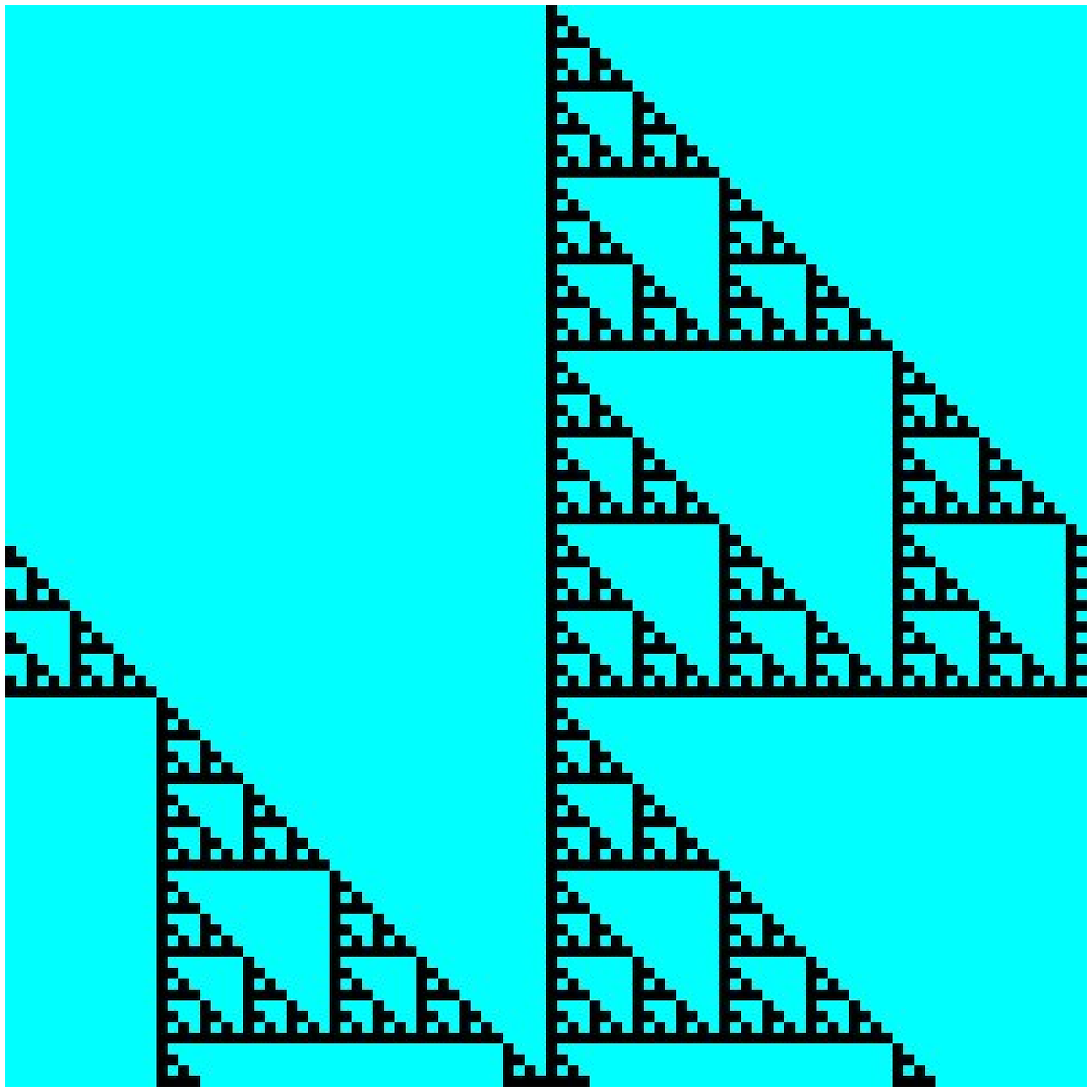}  &   \includegraphics[width=35mm]{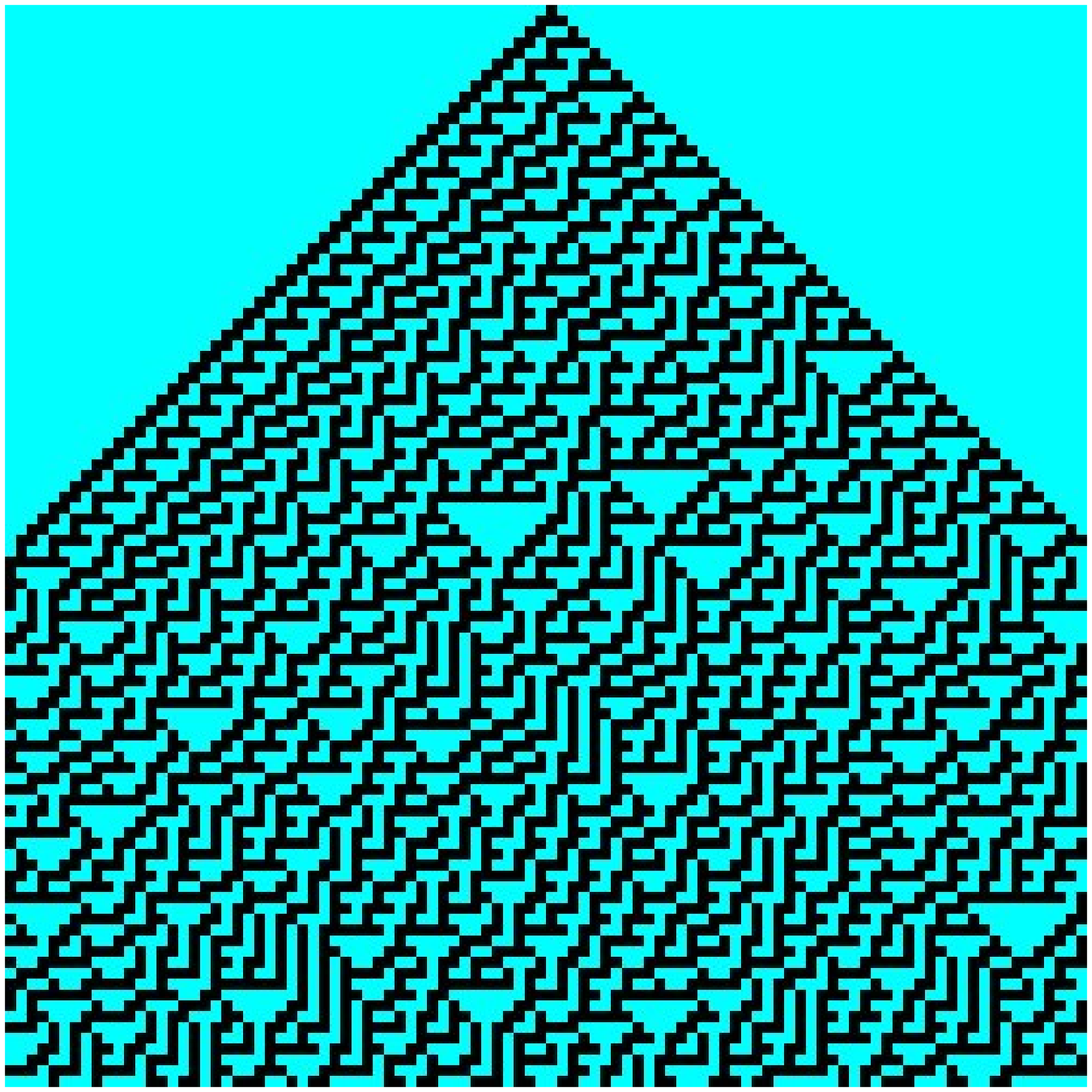} & \includegraphics[width=35mm]{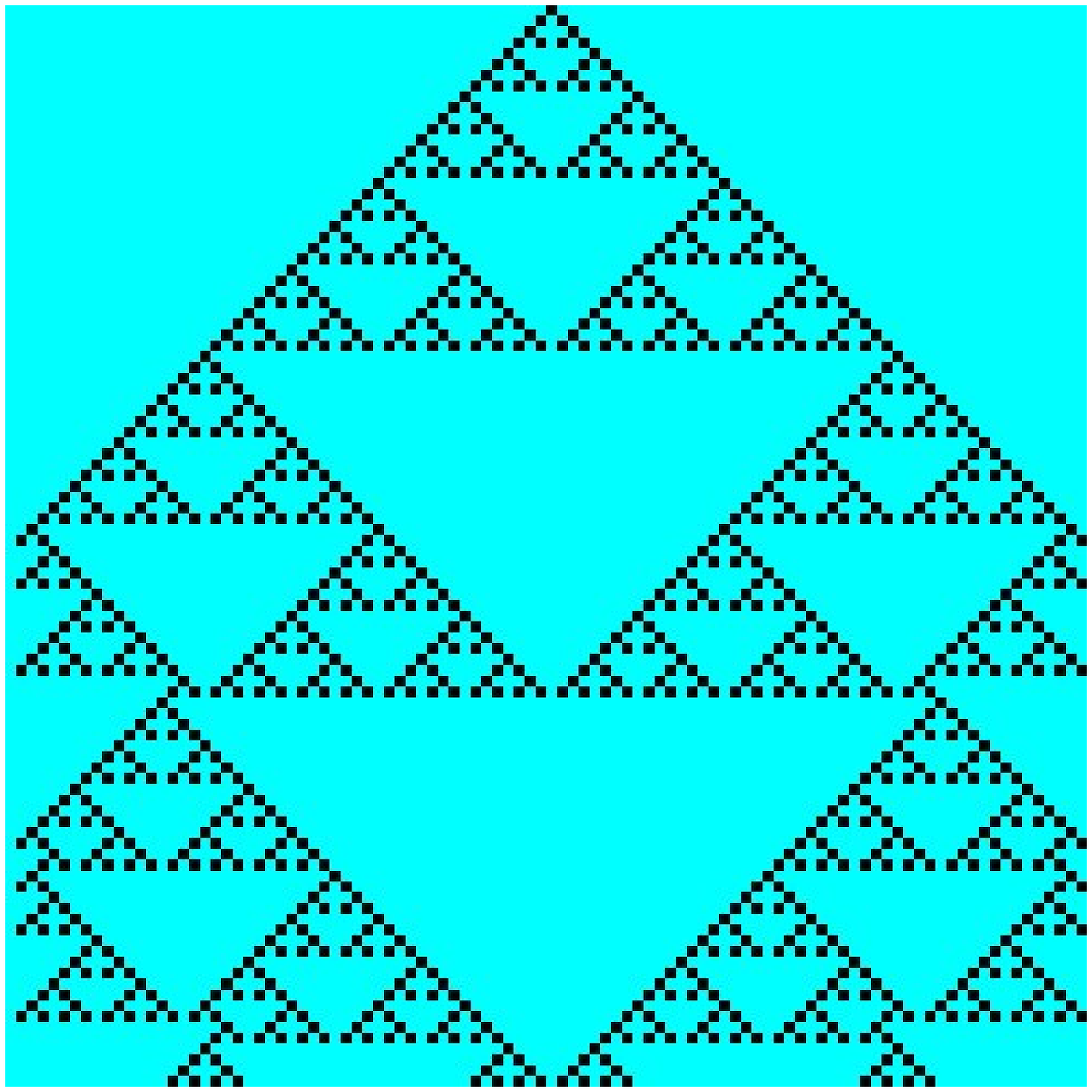}\\

\includegraphics[width=35mm]{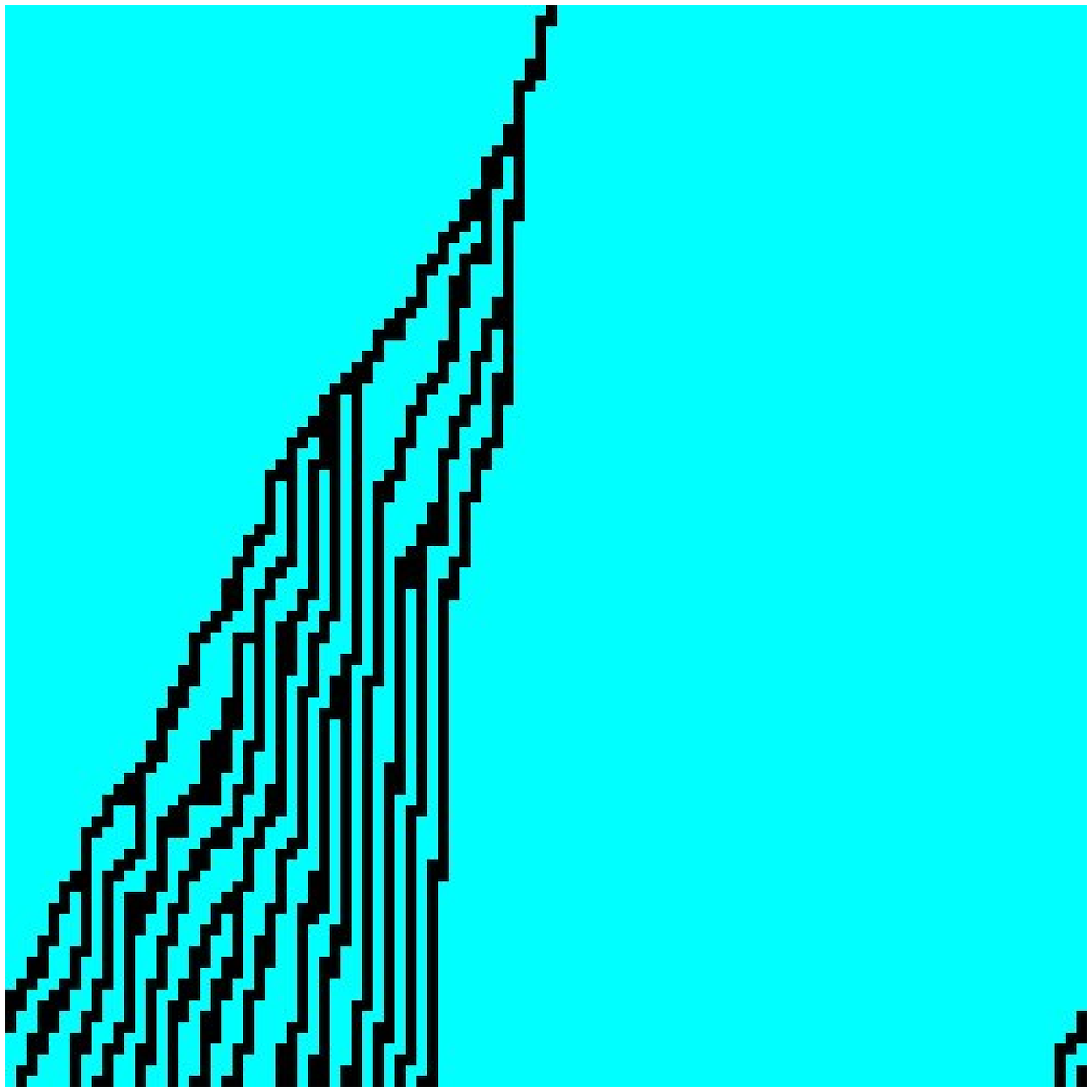} & \includegraphics[width=35mm]{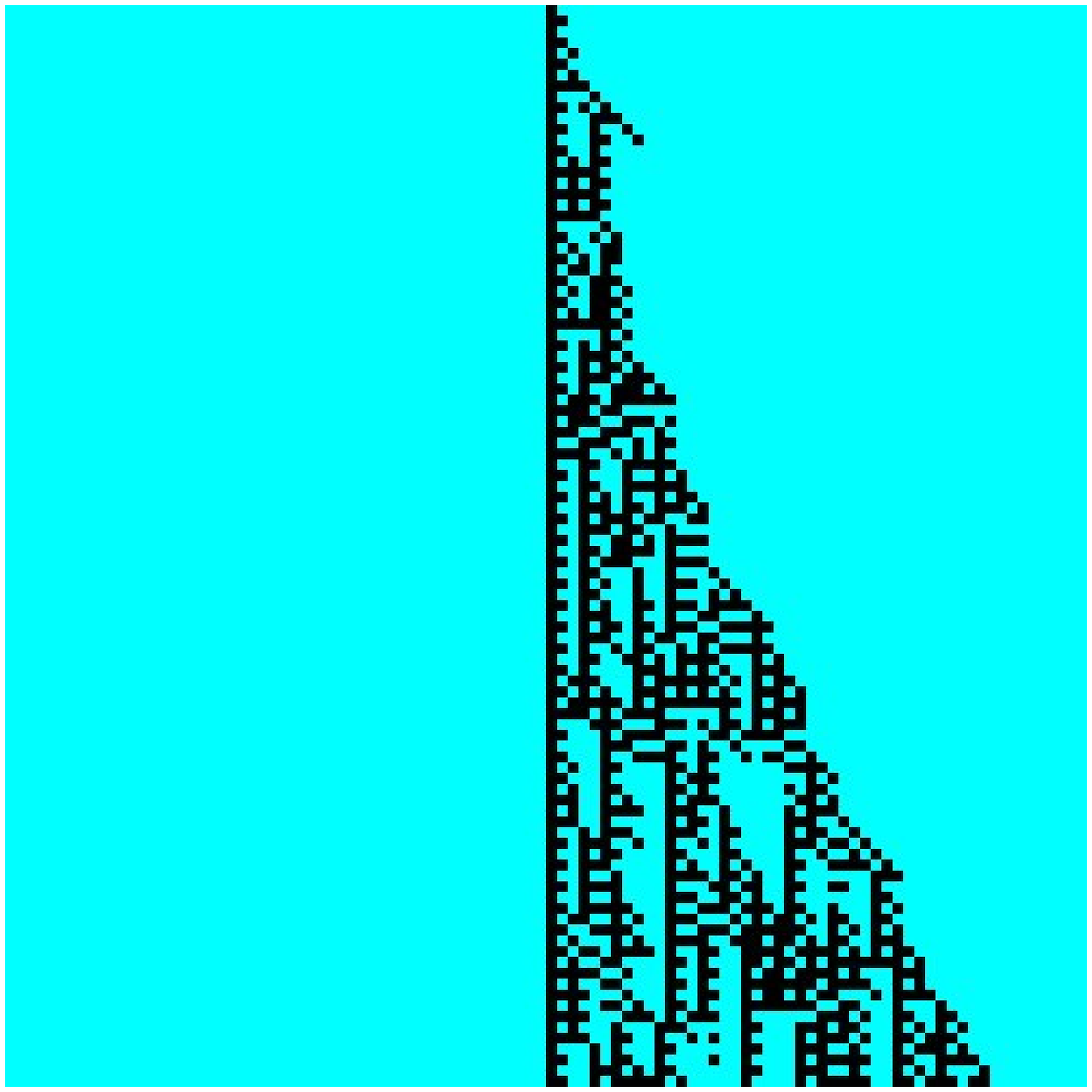}  &   \includegraphics[width=35mm]{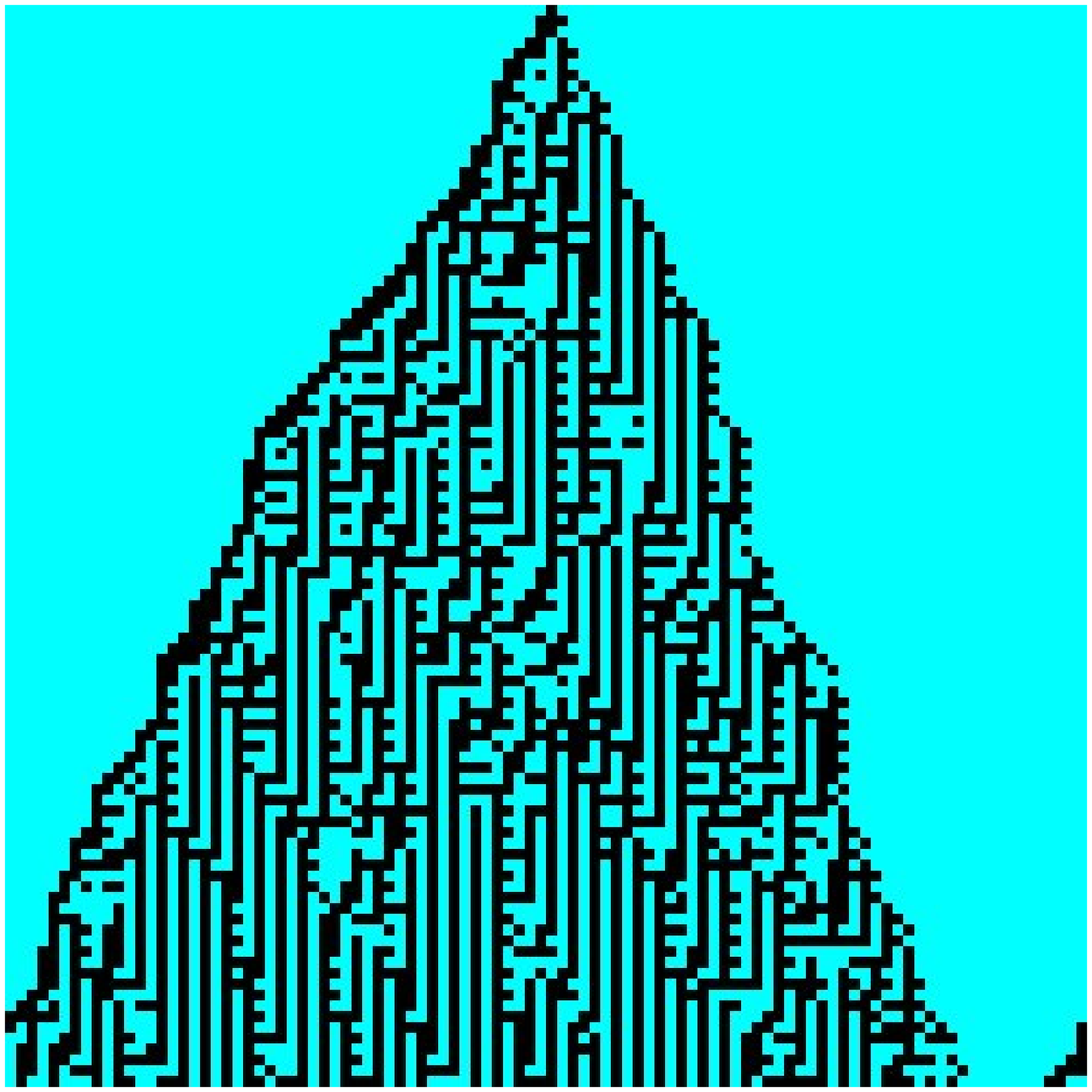} & \includegraphics[width=35mm]{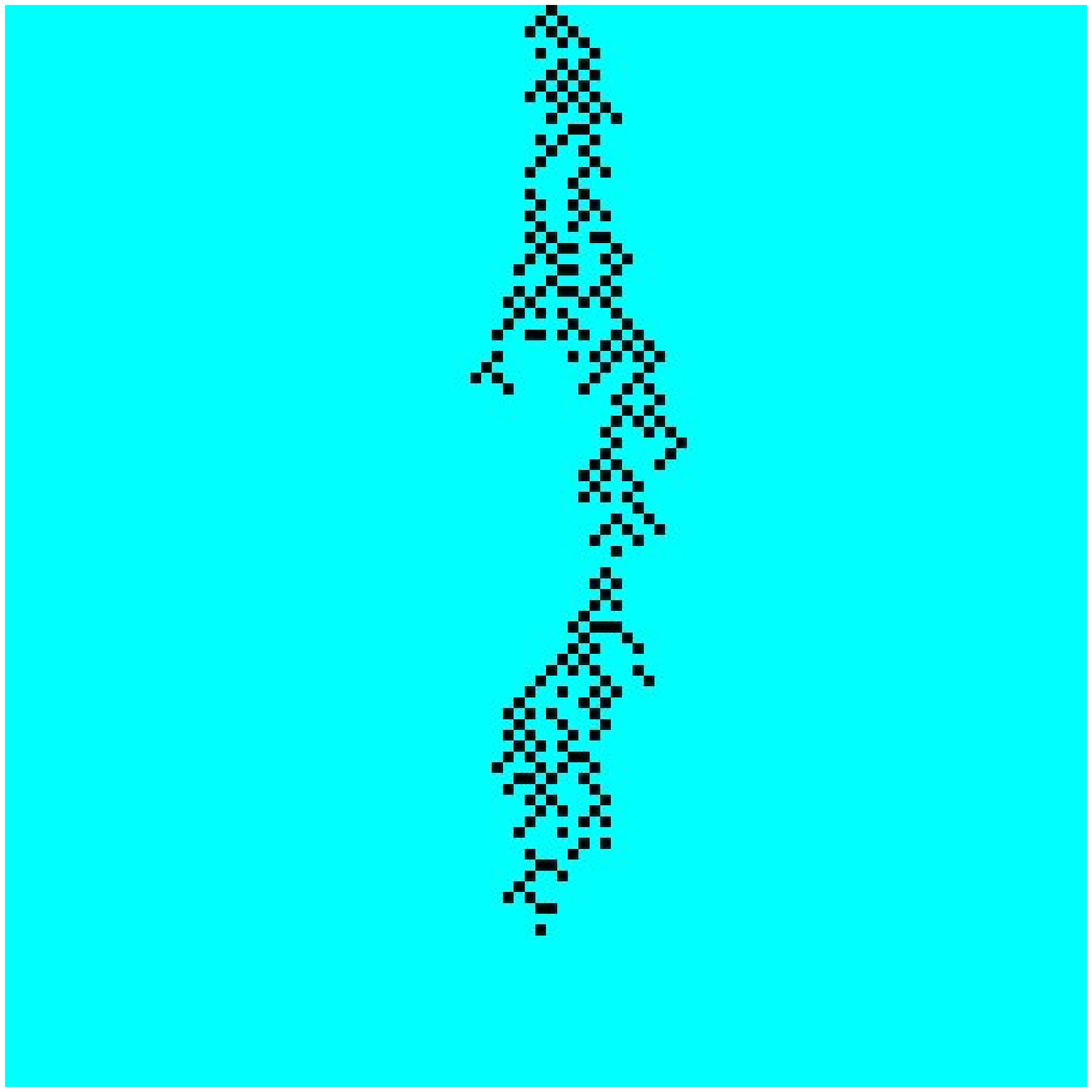}\\

\includegraphics[width=35mm]{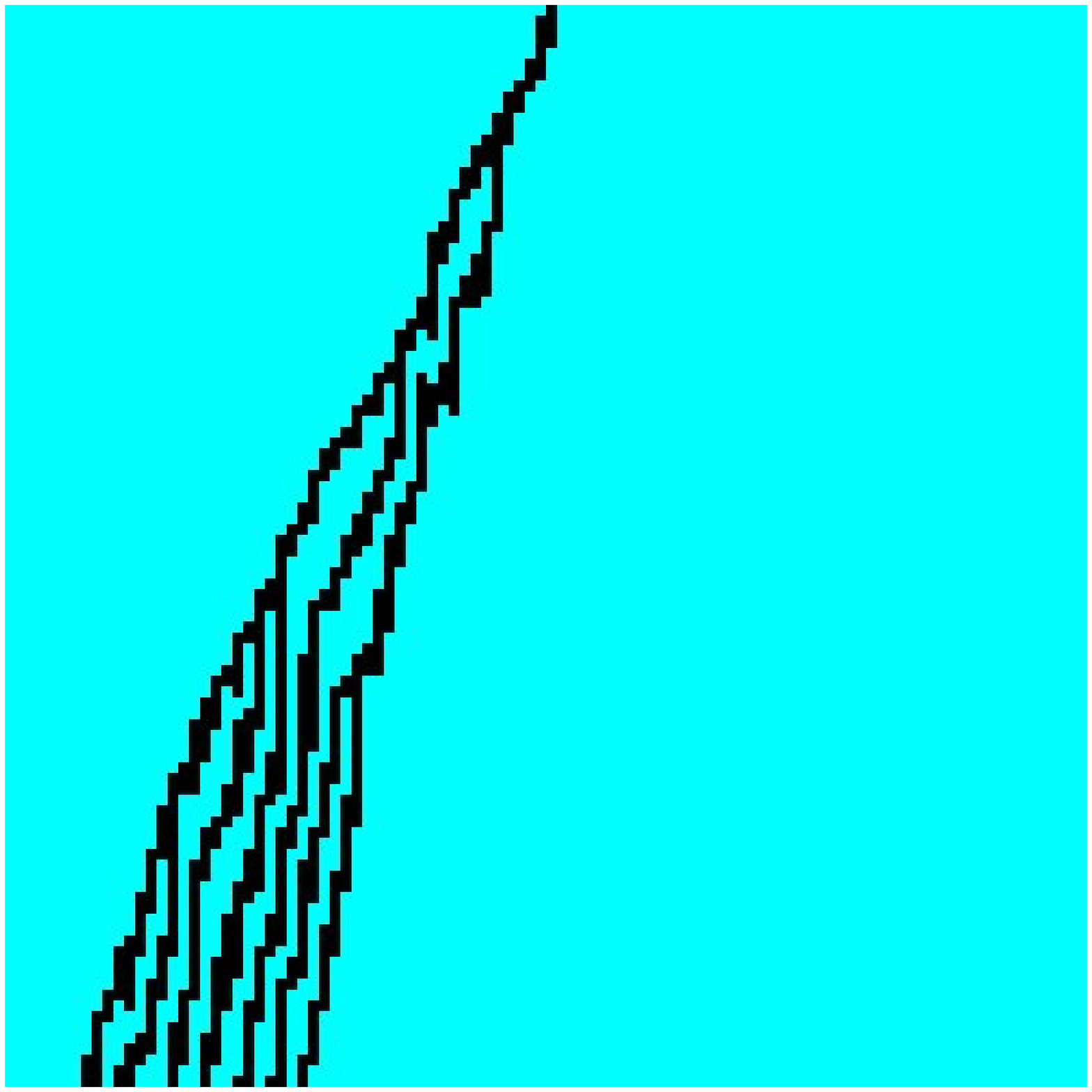} & \includegraphics[width=35mm]{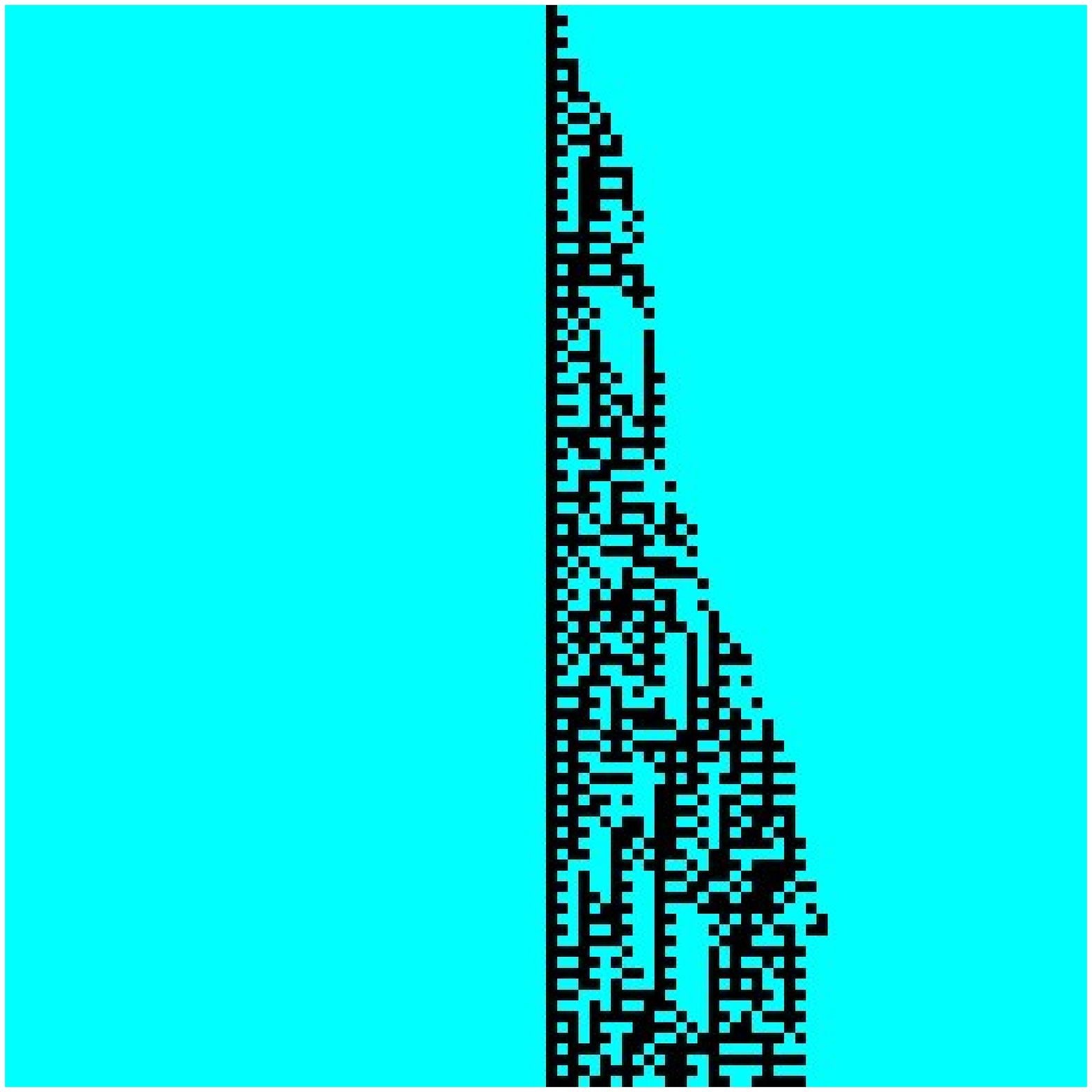}  &   \includegraphics[width=35mm]{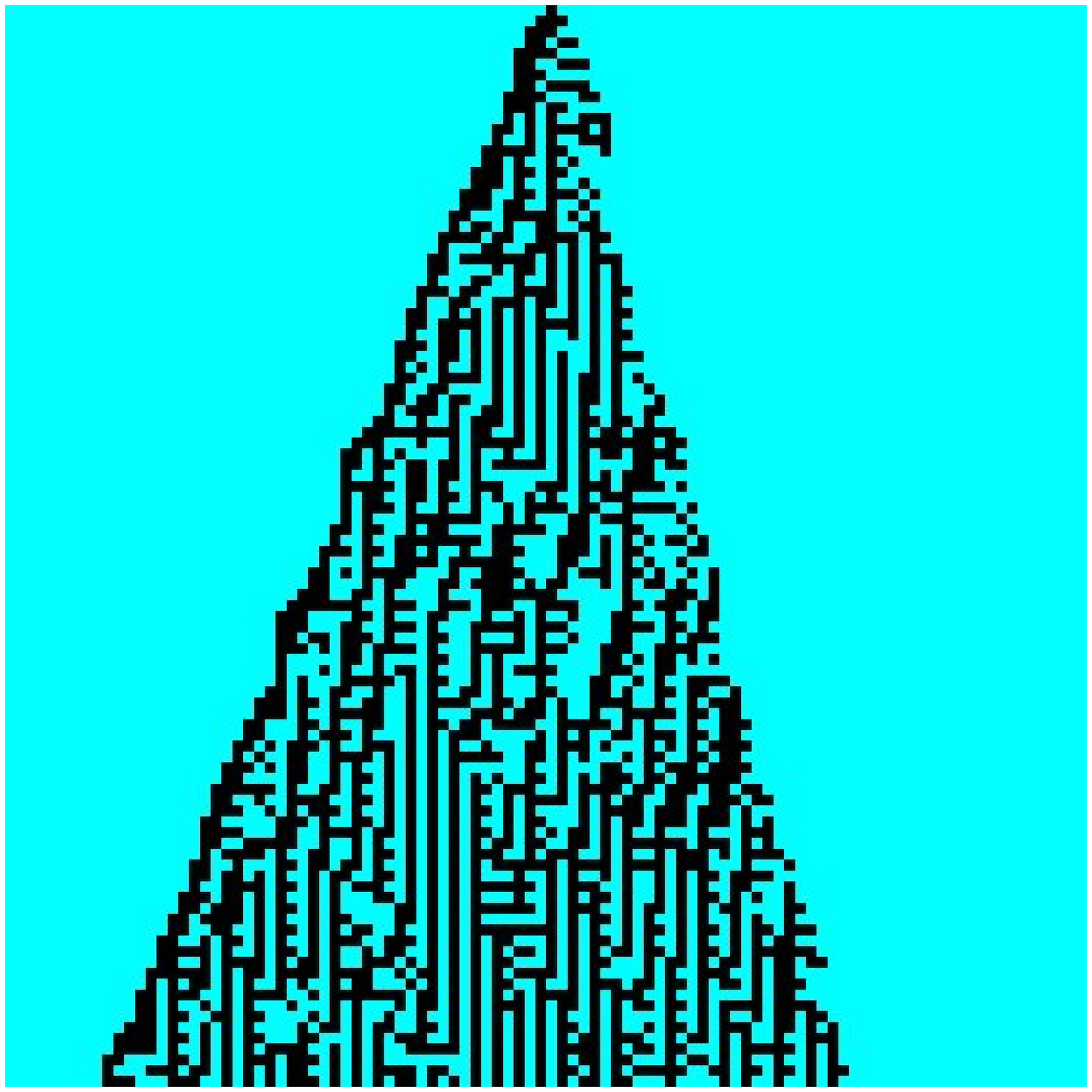} & \includegraphics[width=35mm]{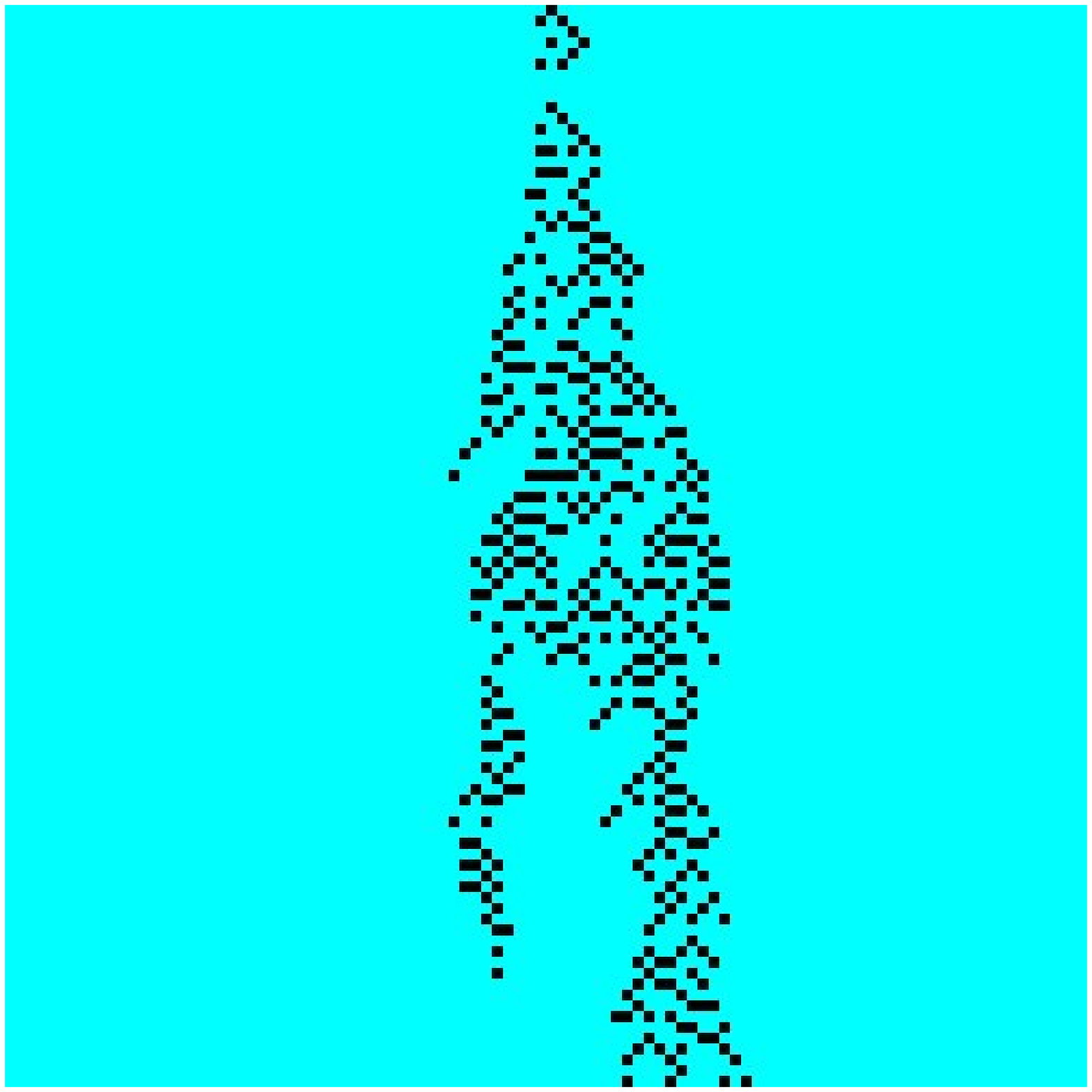}\\
\end{tabular}
\caption{The samples of space time diagrams for the proposed updating schemes - (top) $d$ = $1$, $\iota$ = $0.0$; (middle) $d$ = $1$, $\iota$ = $0.5$; (bottom) $d$ = $2$, $\iota$ = $0.5$.}
\label{space}
\end{figure*}

To model the riot dynamics, we investigates the generative behaviour of the proposed ECA system. During this study, we start with the smallest possible seed pattern as initial configuration where a single cell is in state $1$, i.e. $\langle \cdots 0001000 \cdots \rangle$. From modelling and theoretical research point of view, investigating dynamics of seed patterns starting with single cell in state $1$ is well establish research approach \cite{MARKUS101,Gravner2011,GRAVNER201264}. From the physical implication point of view, the initial seed with a single cell in state $1$ represents the triggering event of riots.

Here, we study the qualitative behaviour of the system starting from a single seed where we need to look at the evolution of the configuration, i.e. space-time diagrams, by inspection over a few time steps. Though this is not a formal method, but this approach can provide a good comparison. Note that, this generative behaviour study does not include $29$ odd ECAs, out of $88$ minimum representative ECAs, which have local transition function $f$($0,0,0$) = $1$. For odd ECAs, an empty background configuration, i.e. $\cdots 000 \cdots$, evolves to $\cdots 111 \cdots$ which is unable to produce the generative behaviour of the system. Therefore, for even $59$ rules, ECA depicts following behaviours - (i) {\em Evolution to zero}: after one time step, the seed cell in state $1$ has vanished; (ii) {\em Constant evolution}: the initial seed remains unchanged during the evolution of the system; (iii) {\em Left evolution}: the seed shifts or grows in the left side; (iv) {\em Right evolution}: the seed shifts or grows in the right; (v) {\em Growth behaviour}: the seed cell develops into a pattern for both left and right side. Fig.~\ref{space} depicts left evolution for ECA $14$, right evolution for ECA $60$ and growth behaviour for ECA $30$. Table ~\ref{clas} shows the classification of ECAs depending on the generative behaviour.

\begin{table} [!htbp]   
\centering 
\begin{tabular}{llllllllll} \hline 
\textit{Evolution to Zero: } & 0 & 8 & 32 & 40 & 72 & 104 & 128 & 136   \\ 
& 160 & 168 & 200 & 232 & & & &\\
\textit{Constant evolution: } & 4 & 12 & 36 & 44 & 76 & 108 & 132 & 140   \\
 & 164 & 172 & 204 &  &  &  &  &    \\
\textit{Left evolution: } & 2 & 6 & 10 & 14 & 34 & 38 & 42 & 46   \\
 & 74 & 78 & 106 & 130 & 134 & 138 & 142 & 162   \\
  & 170 &  &  &  &  &  &  &    \\
\textit{Right evolution: } & 24 & 28 & 56 & 60 & 152 & 156 & 184 &    \\
\textit{Growth behaviour: }& 18 & 22 & 26 & 30 & 50 & 54 & 58 & 90   \\
 & 146 & 150 & 154 & 178 &  &  &  &    \\
\hline
 
\end{tabular}
\caption{Classification of ECA rules depending on the generative behaviour.}
\label{clas}
\end{table}

Here, the target of this simple classification is to identify the ECAs which develop into a pattern for both left and right side, i.e. growth behaviour. Here, we make a sensible simple assumption that the riot propagation affects every neighbour, i.e. both left and right for ECA. Therefore, $12$ ECAs, out of $88$, with growth behaviour are our target for modelling riot dynamics. In this context, note that, Redeker et. al. \cite{MARKUS101} classifies the behaviour of traditional synchronous CA starting from a single seed into - `evolution to zero',`finite growth',`periodic patterns',`Sierpi\`nski patterns' and `complex behaviour' which have no clear equivalence to Wolfram's classes \cite{WOLFRAM19841}. `Evolution to zero' class only shows similarity with this study. Here, the stable structure gets quickly destroyed in the presence of delay and probabilistic loss of information perturbation. As an evidence, in Fig.~\ref{space}, fractal-like Sierpi\`nski patterns \cite{MARKUS101} are destroyed for ECA $18$ under the proposed system. Therefore, now, the target of this study is to identify candidate ECAs from $12$ growth behaviour ECAs to model the riot dynamics.

\subsection{Candidate ECAs for modelling riots}

\begin{figure}
\begin{tabular}{cccc}
  \includegraphics[width=27mm]{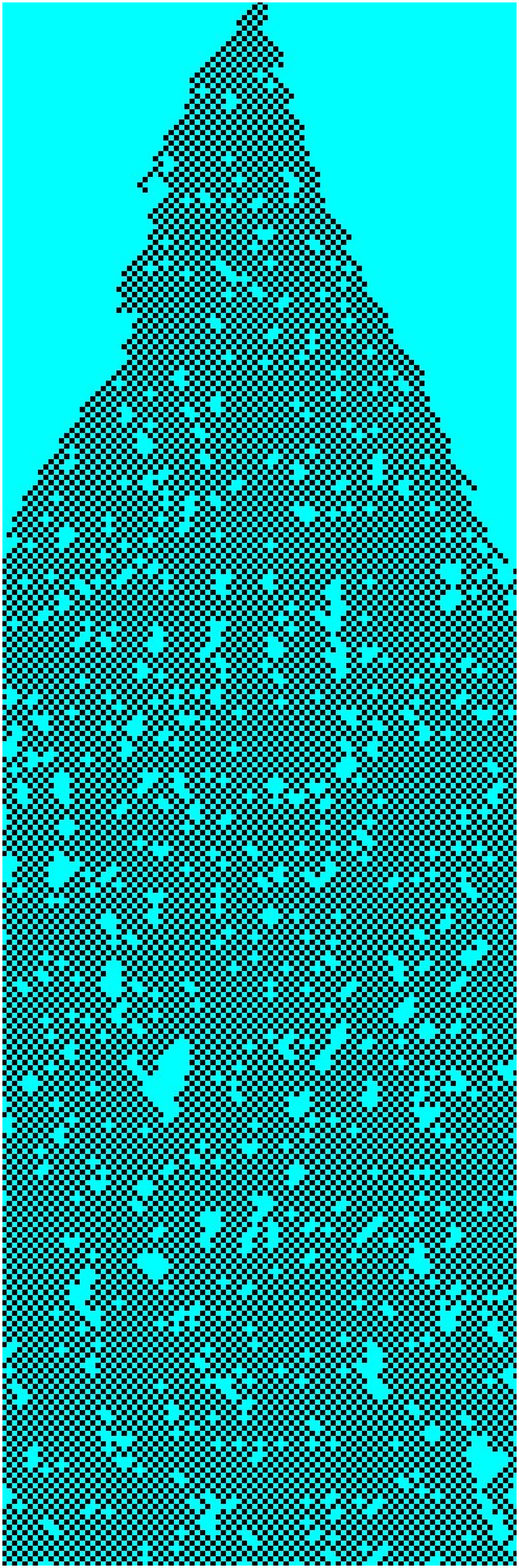} & \includegraphics[width=27mm]{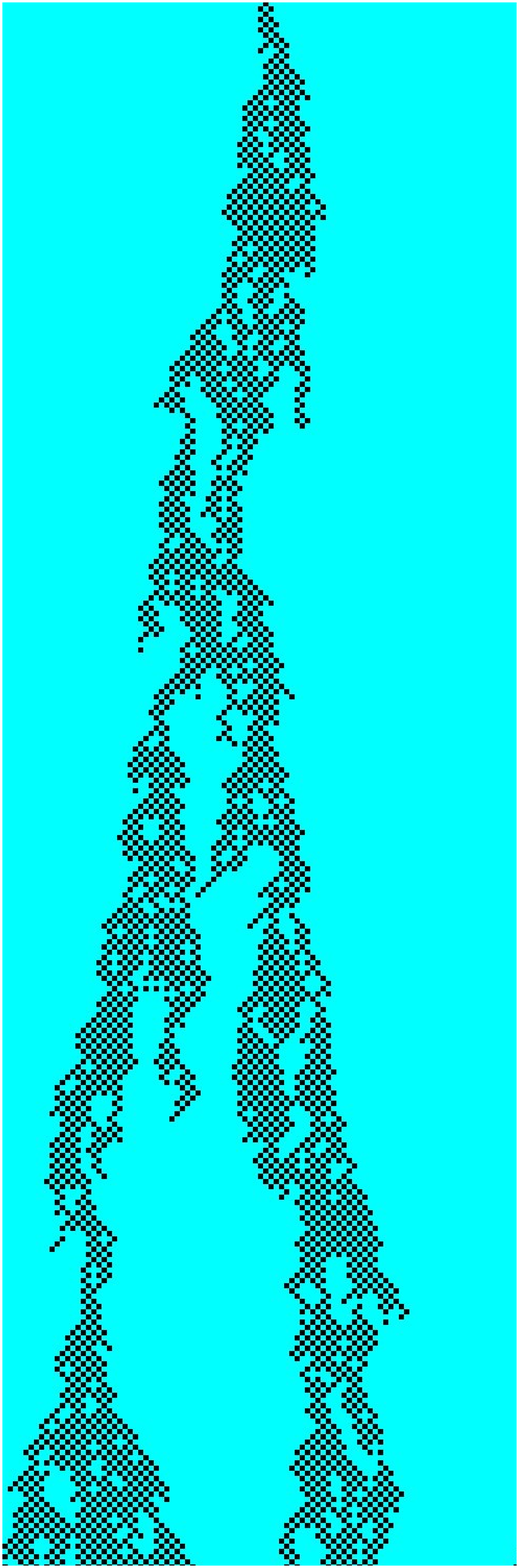}  &   \includegraphics[width=27mm]{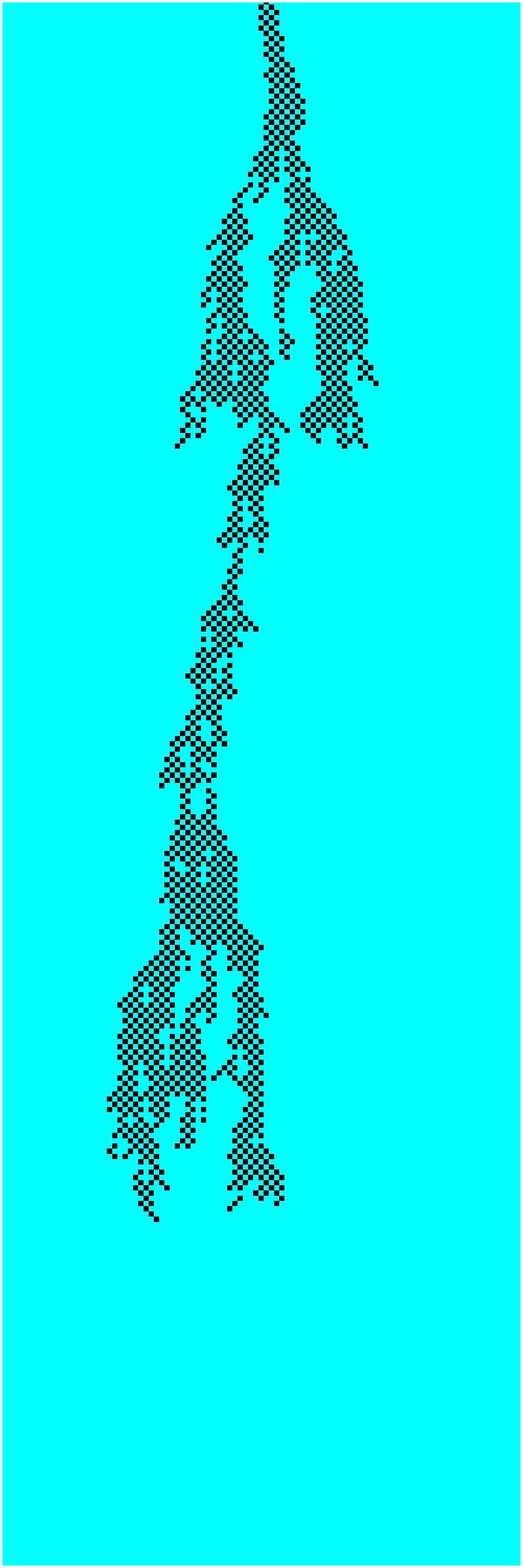}\\
\end{tabular}
\caption{The samples of space time diagrams depicting phase transition - (left) $d$ = $1$, $\iota$ = $0.3$; (middle) $d$ = $1$, $\iota$ = $0.4$; (right) $d$ = $1$, $\iota$ = $0.5$.}
\label{space1}
\end{figure}

Let us assume that the riot dynamics has two phases - spreading phase and diminishing phase. So, we choose $4$ candidate ECAs, out of $12$, which show phase transition under the proposed updating scheme, to model the riot dynamics (as an example, see ECA $18$ in Fig~\ref{space}). For these four ECAs $18$,~$26$,~$50$ and $146$, out of $88$ minimal representative rules, there exists a critical value of probabilistic loss of information perturbation rate which distinguishes the behaviour of the system in two different phases - passive phase (i.e. the system converges to a homogeneous fixed point of all $0$'s) and active phase (i.e. the system oscillates around a non-zero density). As an example for ECA $50$, Fig.~\ref{space1}(left) depicts the active phase for probabilistic loss of information perturbation rate $0.3$ ($\iota$ = $0.3$), however, a phase transition from active to passive phase is observed in Fig.~\ref{space1}(right) where $\iota$ = $0.5$. As physical implication with respect to rioting dynamics, the diminishing phase is not observed without the presence of certain percentage of anti-riot population, i.e. probabilistic loss of information perturbation rate. However, presence of certain percentage of anti-riot population in the society leads to passive phase in the diminishing rioting dynamics. According to \cite{ROY2019600},this phase transition belongs to the directed percolation universality class. Note that, in the literature of rioting dynamics research, the idea of critical threshold was also discussed in \cite{Berestycki11,LANG201412} for understanding level of social tension to start a riot and sufficiently large number of protests to start a revolution respectively.  

\begin{table} [!htbp]   
\centering 
\begin{tabular}{ccccc} \hline 
ECA & $\iota^c_{d=1}$ & $\iota^c_{d=2}$ & $\iota^c_{d=3}$ & $\iota^c_{d=4}$ \\ \hline

$18$ & $.27$ & $.38$ & $.46$ & $.48$ \\
$26$ & $.51$ & $.54$ & $.63$ & $.67$ \\
$50$ & $.48$ & $.50$ & $.53$ & $.55$ \\
$146$ & $.32$ & $.43$ & $.46$ & $.50$ \\ \hline
 
\end{tabular}
\caption{The critical value for phase transition of ECA $18$,~$26$,~$50$ and $146$.}
\label{cri}
\end{table}

\begin{figure*}
\begin{tabular}{cccc}
  \includegraphics[width=45mm]{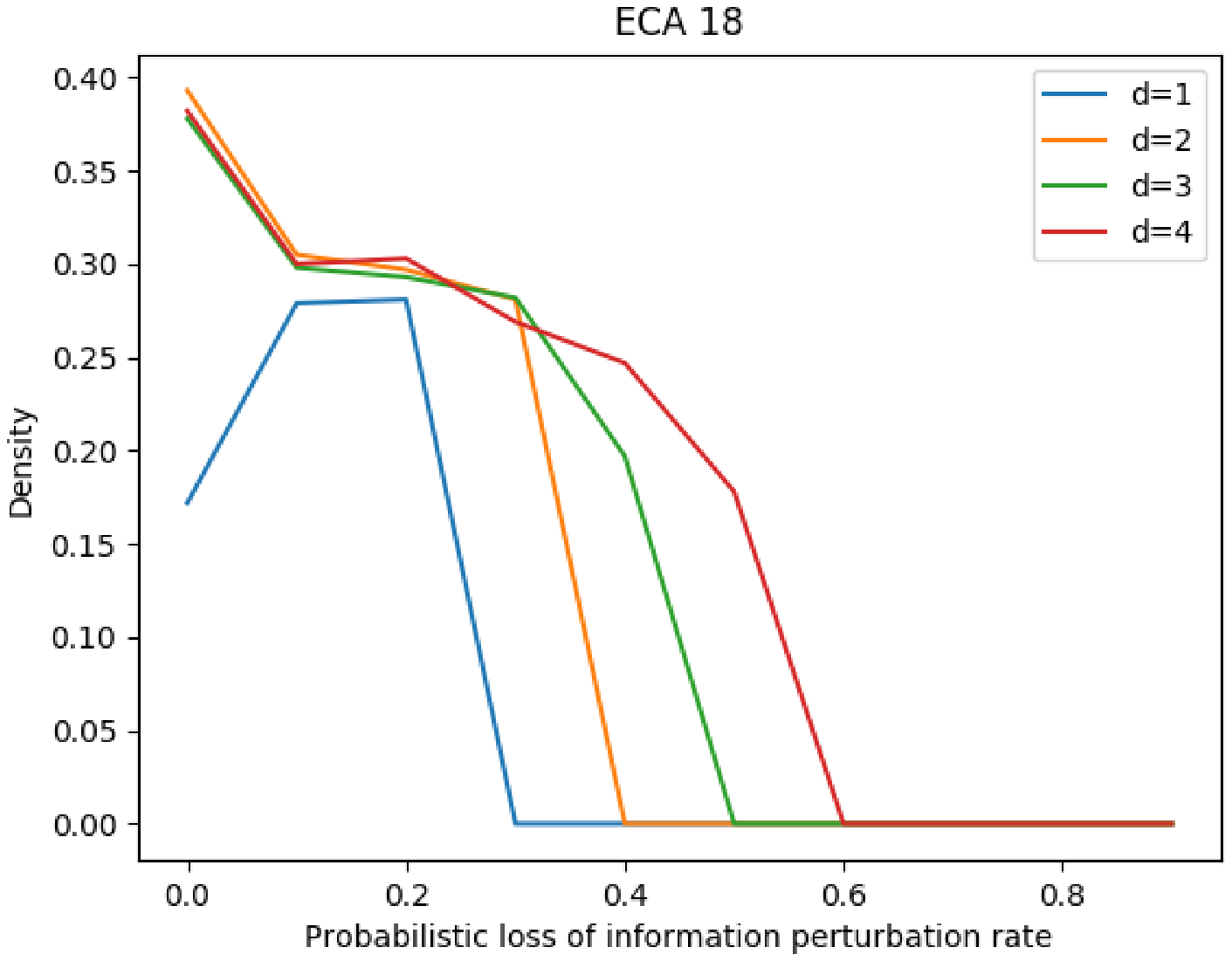} & \includegraphics[width=45mm]{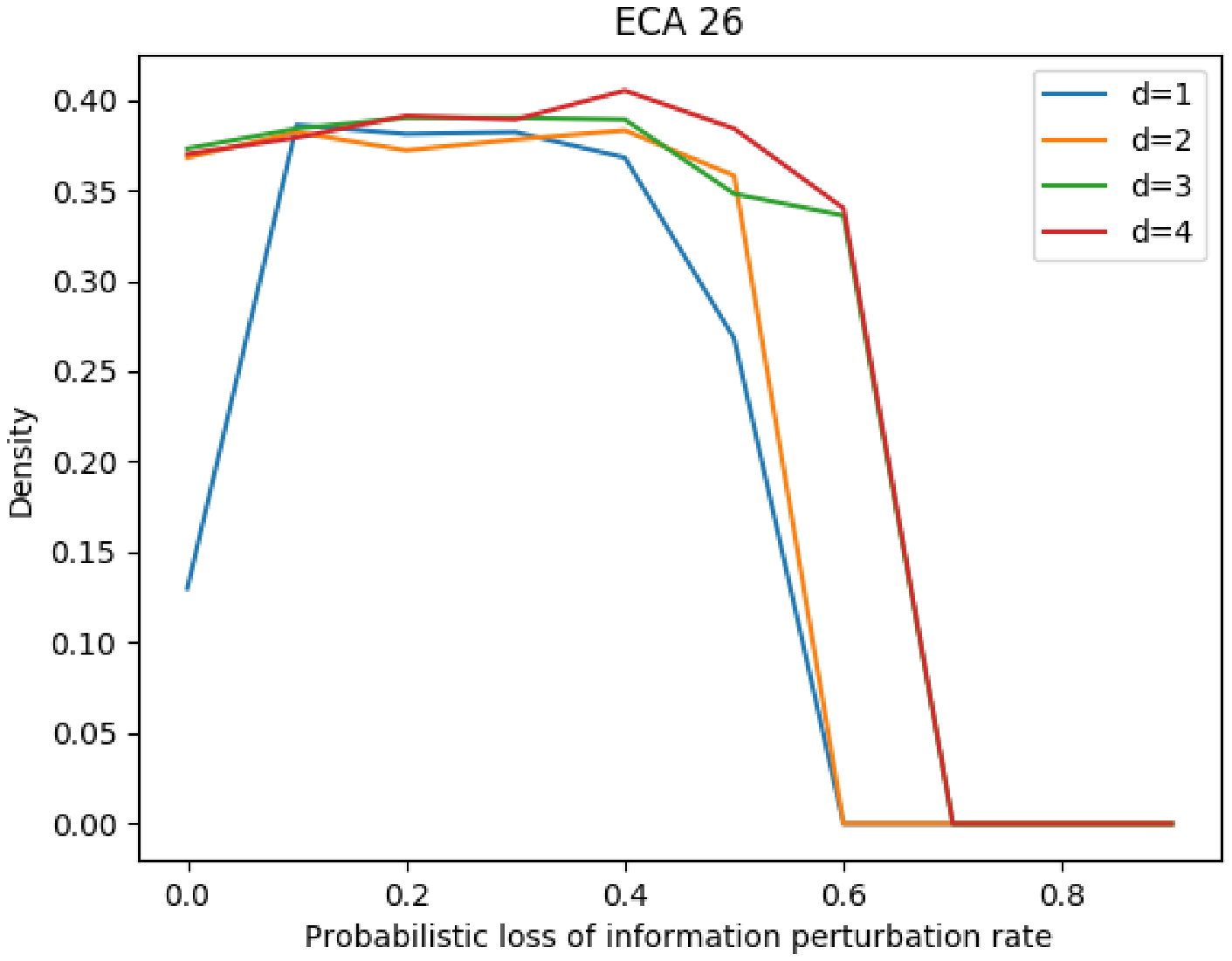}  &   \includegraphics[width=45mm]{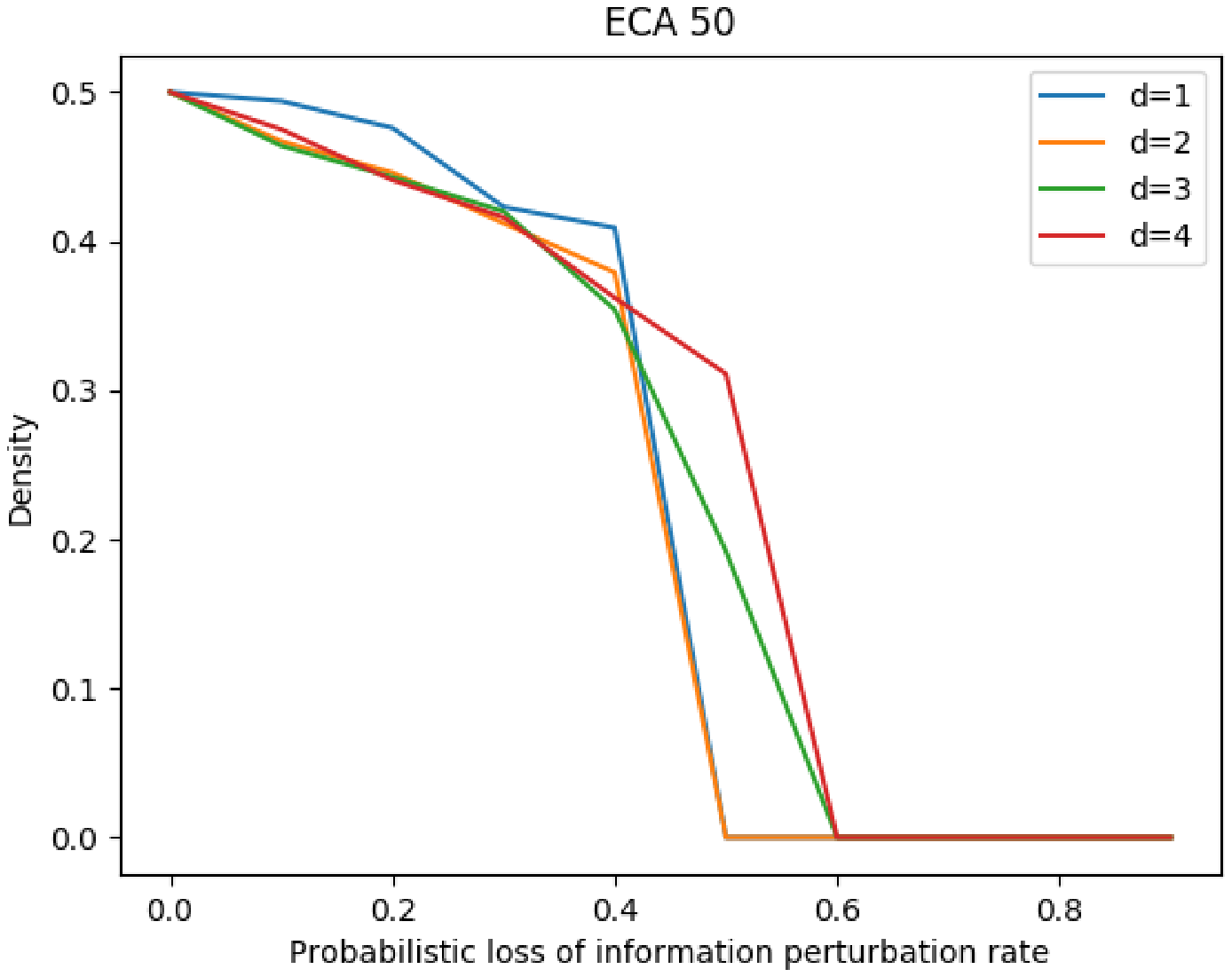} & \includegraphics[width=45mm]{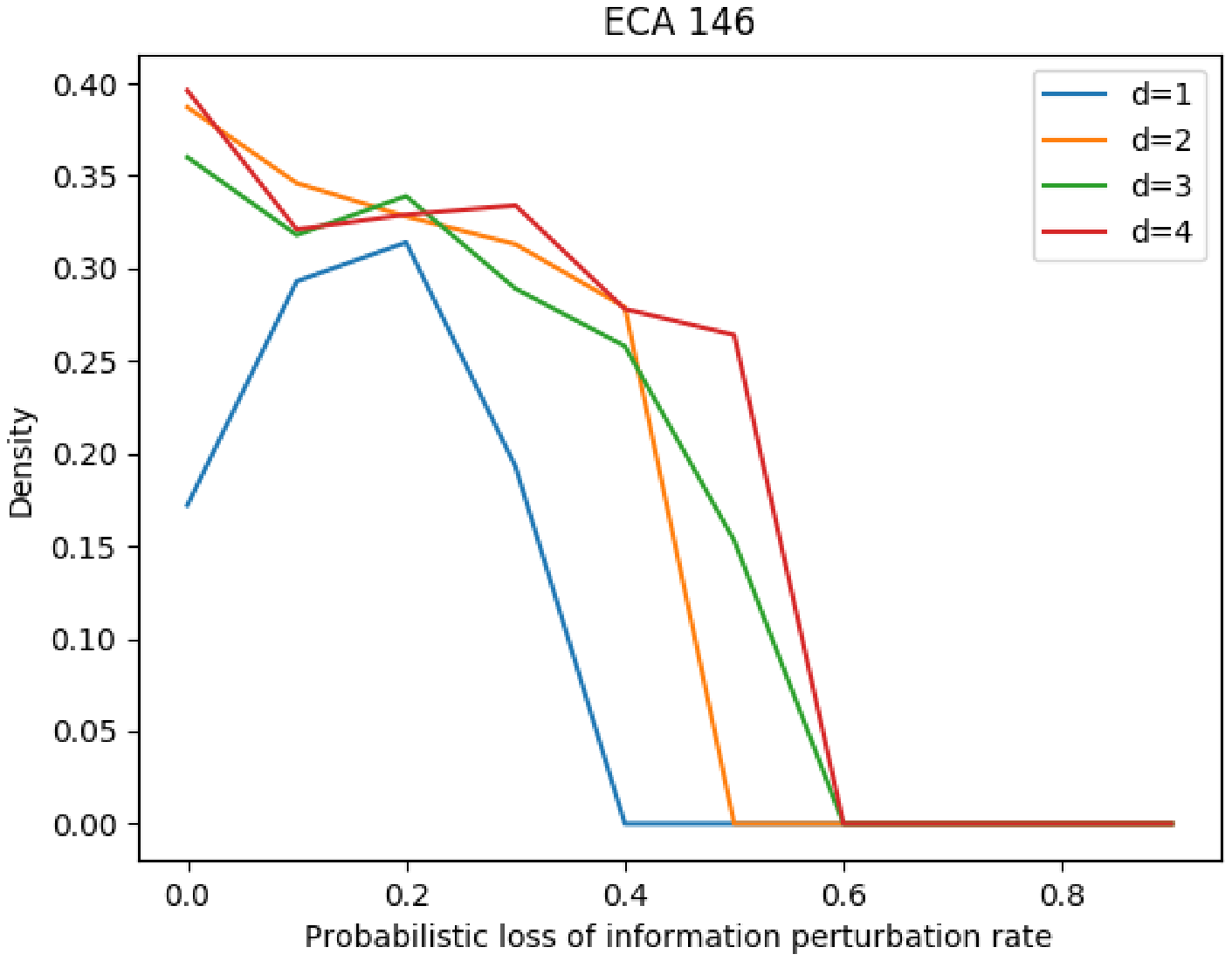}\\
\end{tabular}
\caption{The plot shows the profile of density parameter as a function of the probabilistic loss of information perturbation rate with a fixed d parameter for ECA rules.}
\label{plot}
\end{figure*}

Now, to understand the quantitative behaviour of these candidate rules, we let the system evolve through $2000$ time steps and average the density parameter value for $100$ time steps. Note that, for a configuration $x \in \mathcal{S}^\mathcal{L}$, the density can be defined as $d$($x$) = $\#_1x$/$\mid$x$\mid$, where $\#_1x$ is the number of $1$'s in the actualstate for the configuration and $\mid$x$\mid$ is the size of the configuration. Fig.~\ref{plot} shows the plot of the profile of density parameter starting from a single `$1$' seed as a function of the probabilistic loss of information perturbation rate with a fixed $d$ parameter for ECA rules which depicts phase transition behaviour. Table~\ref{cri} depicts the critical value of probabilistic loss of information perturbation rate for phase transition associated with these ECA rules where $\iota^c_{d=k}$ indicates the critical value with $d$ parameter value $k$. Note that, the critical value for phase transition increases when the updating scheme is also associated with delay perturbation, see Table~\ref{cri} for evidence. Moreover, the critical value of probabilistic loss of information perturbation rate for phase transition proportionally increases with increasing value of delay. Table~\ref{cri} justifies that the diminishing phase of riot needs more percentage of anti-riot population in the presence of sociological factor delay. Note that, this phase transition result is not observed for only delay perturbation updating scheme. To sum up, ECAs $18$,$26$,$50$,$146$ are the final candidate rules for modelling rioting dynamics. Therefore, now, the target is to identify the best candidate rule among those ECAs for validation of Baduria riot dynamics. In this scenario, the next section depicts the case study comprising Baduria riot's dataset.

\section{Baduria riot and the proposed system}

\subsection{Baduria riot dataset}
Attracting nationwide media attention, Baduria riot is the most well-exposed among recent Bengal's riot events \cite{indiatoday2016,hindustantimes2018}. The triggering event of Baduria riot took place after a social media religious post by a $17$-year old student in a village Baduria of West Bengal on $2^{nd}$ July, 2017. This social media post was seen as objectionable and went viral in Baduria. Starting with this, violent clashes were triggered between the two communities of Baduria, Basirhat, Swarupnagar, Deganga, Taki and Haroa sub-division.

\begin{figure}[!htbp] 
\centering 
\includegraphics[width=3.2in]{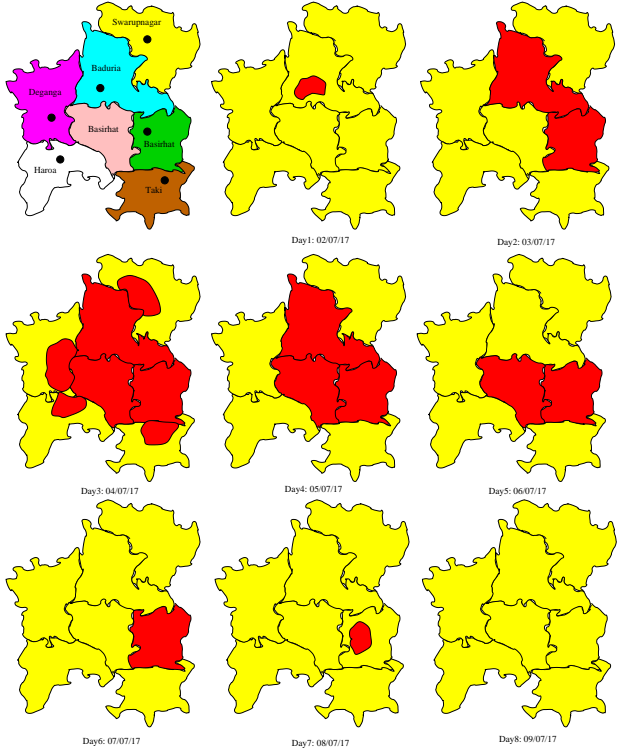} 
\caption{Graphical riot propagation dynamics of Baduria riot.}
\label{p1}
\end{figure}

Here, we base our analysis here on reported incidents in media reports during Baduria rioting time. The authenticity of media report data is cross verified with field study. We extract riot like events, as examples `attack on police vehicles', `serious injuries', `group clashed' etc., from $20$ media reports 
\cite{bbc2017,financialexpress2017,news182017,EXPRESS17,NDTV,hina1,firstpost,hina2,
hina3,hina4,hina5,hina6,hina7,hina8,hina9,hina10,hina11,hina12,hina13,hina14} to build the data set. In the literature, the traditional methodology for quantifying the rioting activity is to study the daily crime reports of police data for analysing the riot dynamics \cite{Laurent211,Davies2013211}. However, this methodology suffers due to lack of data on rioting event geographically located in third-world country. Note that, arrest records, though available, do not indicate communal riot as explicit reason for the arrest.

\begin{figure*}
\begin{tabular}{cccc}
 \includegraphics[width=45mm]{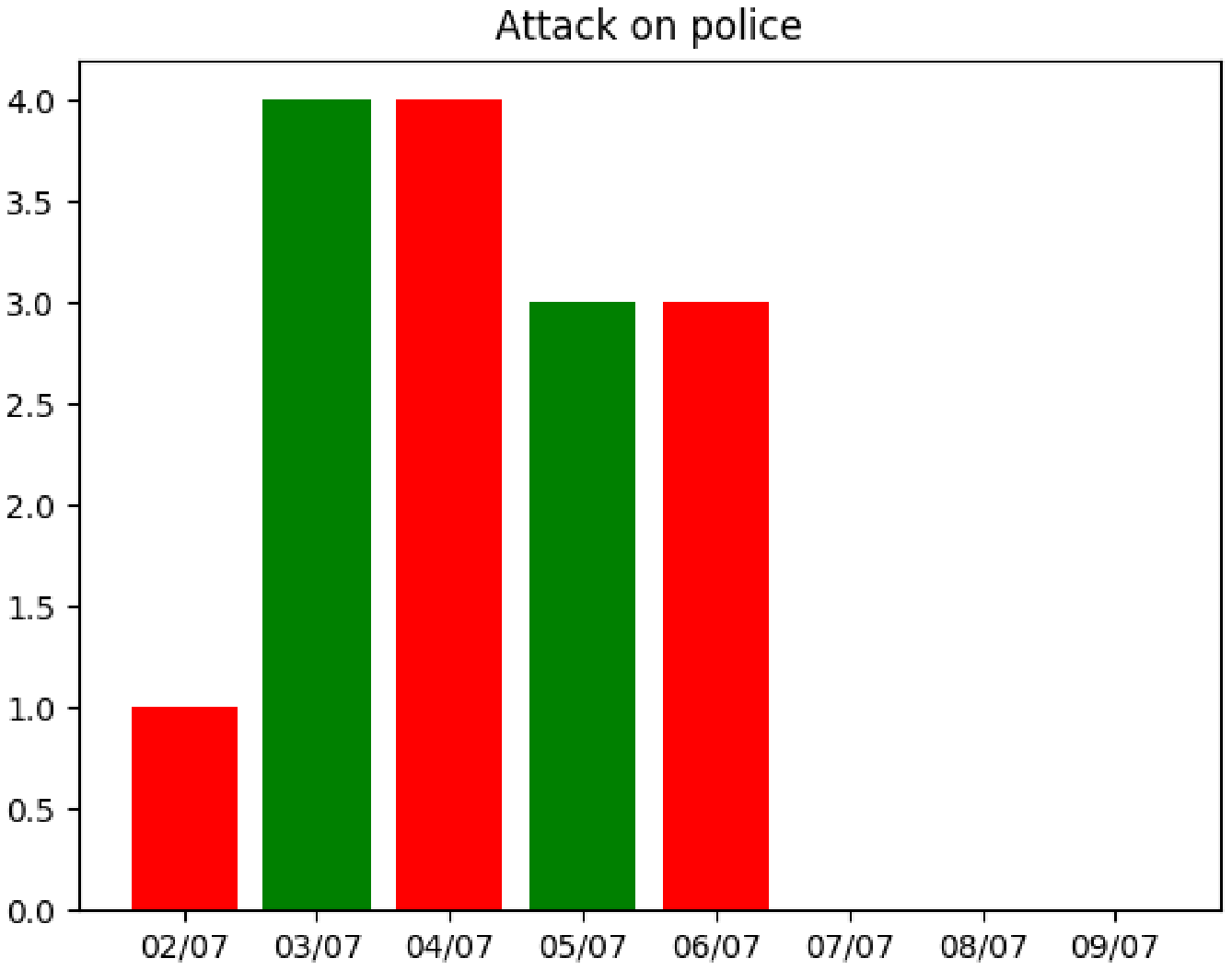} & \includegraphics[width=45mm]{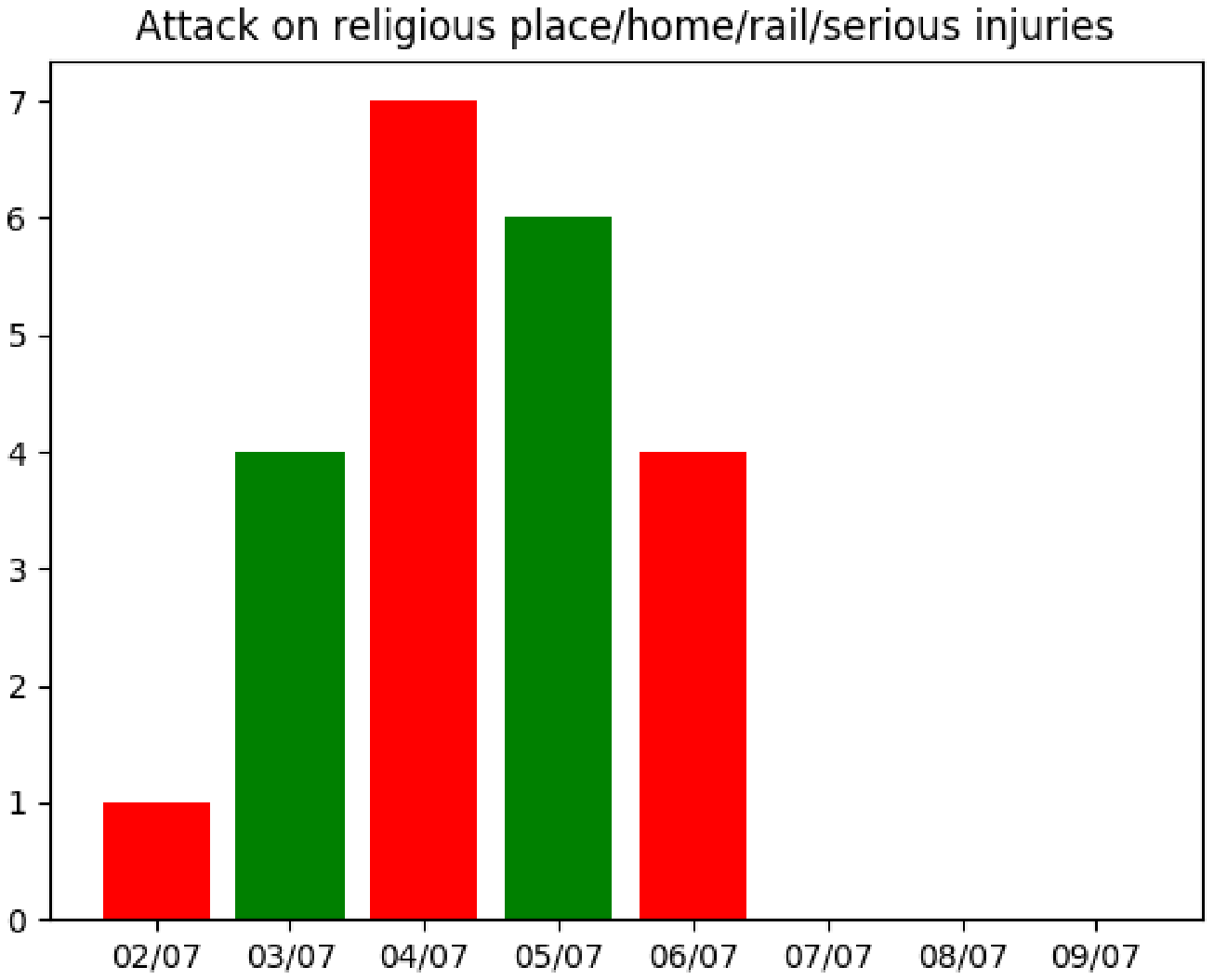}  &   \includegraphics[width=45mm]{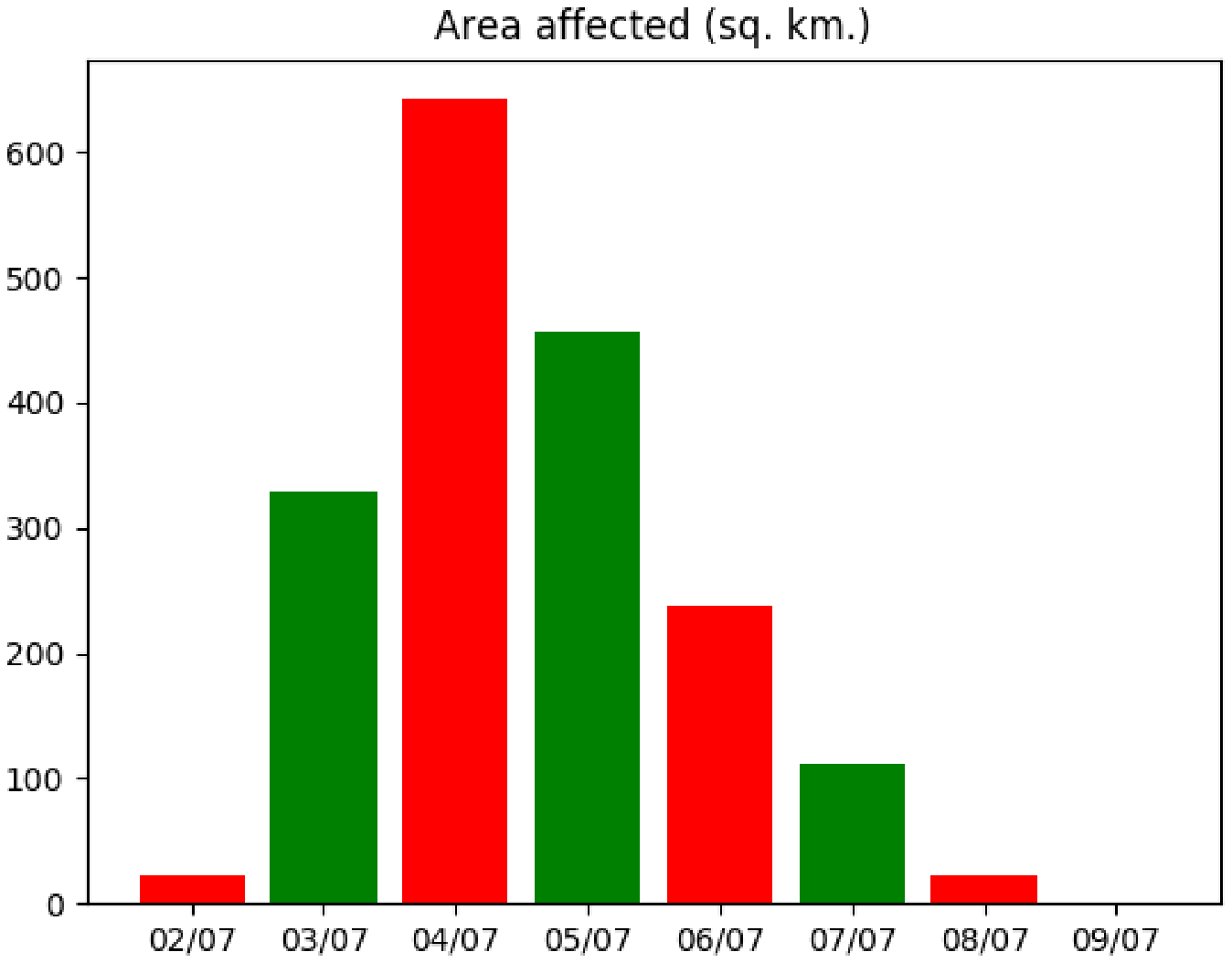} & \includegraphics[width=45mm]{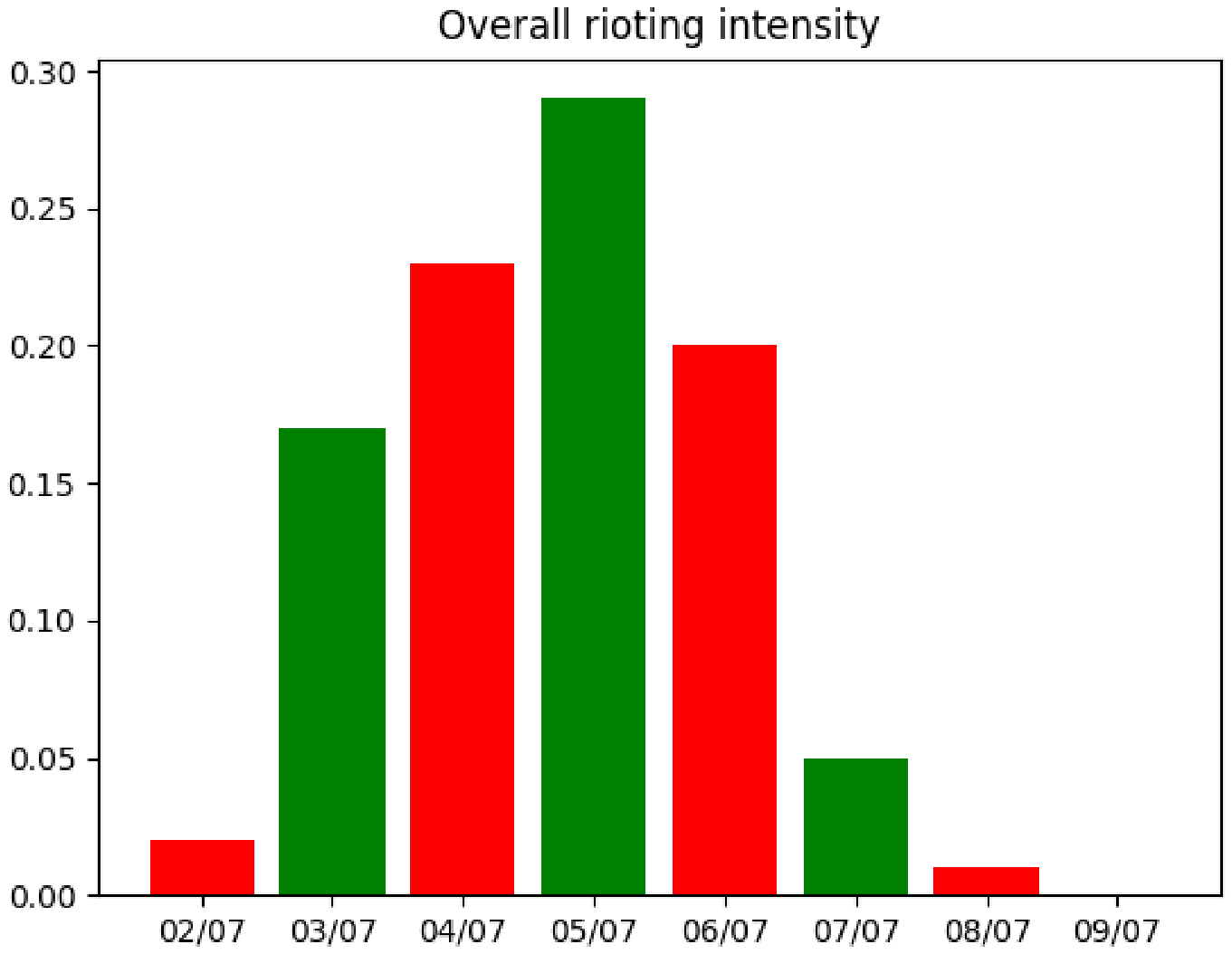}\\
(a) & (b) & (c)& (d) \\
\end{tabular}
\caption{The plot shows quantified rioting activity for every day from July $2$ to July $9$, $2017$ of ($a$) attack on police (vehicles,stations,persons); ($b$) attack on religious place, home, rail line and serious injuries; ($c$) affected area in riot (by sq. km.); ($d$) normalized overall intensity.}
\label{p2}
\end{figure*}

Here, we adopt two simple methodology for quantifying the rioting activity: Firstly, we define as a single event any rioting-like act, as listed in the media reports after ambiguity checking, depending on its intensity. Thus, `rail blockades at $3$ places' counts as $3$ events, `three police vehicles have been torched' indicates $3$ riot like events. We thus get a dataset composed of number of riot like events for every day from July $2$ to July $9$, $2017$. Secondly, we quantify rioting activity from area (by sq. km.) affected by riot on a day to day basis from media reports. Fig.~\ref{p1} reflects the spatial propagation of riot dynamics. It is not possible to quantify some important riot events, like (number of) group clashes, road blockades, market/shop/school closure, from media reports. Therefore, we quantify those rioting events by area affected in riots day-wise. Figure~\ref{p2} shows number of attack events on (a) police (vehicles, stations, persons); (b) religious place, home, rail line and serious injuries; (c) affected area (in sq. km.) over the of time course.

Now, we calculate the summation of percentage of intensity per rioting day, out of total intensity, for the rioting event datasets of Fig.~\ref{p2}(a),(b),(c). Hereafter, the percentage of intensity per rioting day of this summarized intensity indicates the {\em normalized overall intensity} of Baduria riot which is reflected in    Figure~\ref{p2}(d). Here, we work with this normalized overall intensity to understand rioting dynamics, which shows simple growth (up) and shrink (down) dynamics. Note that, this simple up and down dynamics, without any rebound, was also observed for $2005$ French riots \cite{Laurent211} and US ethnic riots \cite{9989878}. With a contradiction, the dynamics with up and sudden down (i.e. rise for four days and suddenly down on fifth day) was found in $2011$ London riots \cite{Davies2013211,Berestycki11}.

On a different internal dynamics perspective, the propagation dynamics of Baduria riot depends on the local communication of violent events 
\footnote{in some village there were incidents of bike burning of Hindu youth. On the spread of this news to some other village, Muslim youths in that village were beaten up. In this way of local communication the riot began to spread. Name: Hindu Salesman; Address: Vill- Choto Jirat, Basirhat; Interview date, time: 20th October, 2019, 1:30pm at Itinda Ghat(near Icchamati River)} 
and local rumour propagation 
\footnote{There were random rumours of attacks the madrasa or temples in the nearby villages. These rumours were propagating the riots. Name: Jallaluddin; Address: Bhabla halt, Basirhat; Occupation: Cycle garage owner; Interview date, time: 20th October, 2019, 3:00pm at Bhabla halt Station (Muslim area)},
\footnote{There were rumours about the attacks upon mandir-masjid but on his verification he found that there were no such incidents. Name: Tapas Ghosh; Address: Paikpara, Basirhat; Occupation: Shopkeeper; Interview date, time: 20th October, 2019, 4:00pm at Paikpara (Muslim area)}. However, during the riot event a parallel journey of religious harmony is also reported in media \cite{financialexpress2017,news182017} which finally converted the dynamics of the event as an early diminishing riot event. The field data also reflects this evidence 
\footnote{As he said a sick Muslim-youth along with his old mother were found sitting near the road. Out of riot fear, they were clueless about what to do and how to reach home. The Muslim-youth was discharged from hospital and was seen carrying saline bottles along with him. The Hindu salesman felt very sympathetic towards the Muslim-youth, he took the Muslim-youth to his house to provide him shelter. Due to this attitude the Hindu-neighbour of the Hindu-salesman at first got angry but later on they understood the situation. Name: Hindu Salesman; Address: Vill- Choto Jirat, Basirhat; Interview date, time: 20th October, 2019, 1:30pm at Itinda Ghat(near Icchamati River)},
\footnote{After the attack a Muslim doctor generously gave Rs. 5000/- to a poor Hindu shopkeeper whose shop was attacked. Name: Tapas Ghosh; Address: Paikpara, Basirhat; Occupation: Shopkeeper; Interview date, time: 20th October, 2019, 4:00pm at Paikpara (Muslim area)}. 

In this context, anti-riot population of society does not participate in this rumour propagation. Moreover, they play an important role in the early diminishing dynamics of the riots. Here, the term `anti-riot' population is composed of following population - firstly, the secular population of society \cite{financialexpress2017,news182017}; secondly, not `purely' secular population, however, due to the economical dependency they play an anti-riot role during the riot time
\footnote{There exists a dependency of Hindu-Muslim in the area, it is seen that Muslim workers work under Hindu owner and vice-versa, and also seen that they work together as workers, that is the main reason of early convergence of the riot. Name: Sambhu Seth; Address: Trimohini, Basirhat; Occupation: Owner of Private Eye Hospital; Interview date, time: 20th October, 2019, 12:00pm}.

\subsection{Verification}

\begin{figure}[!htbp] 
\centering 
\includegraphics[width=3.0in]{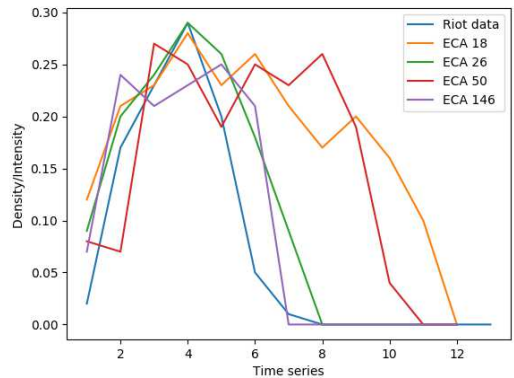} 
\caption{The plot compares normalized overall intensity of Baduria riot and normalized density of ECA rules as a function of time series. Here, $d = 1$ and $\iota = \iota^c$.}
\label{c1}
\end{figure}

\begin{figure*}
\begin{tabular}{cccc}
\includegraphics[width=45mm]{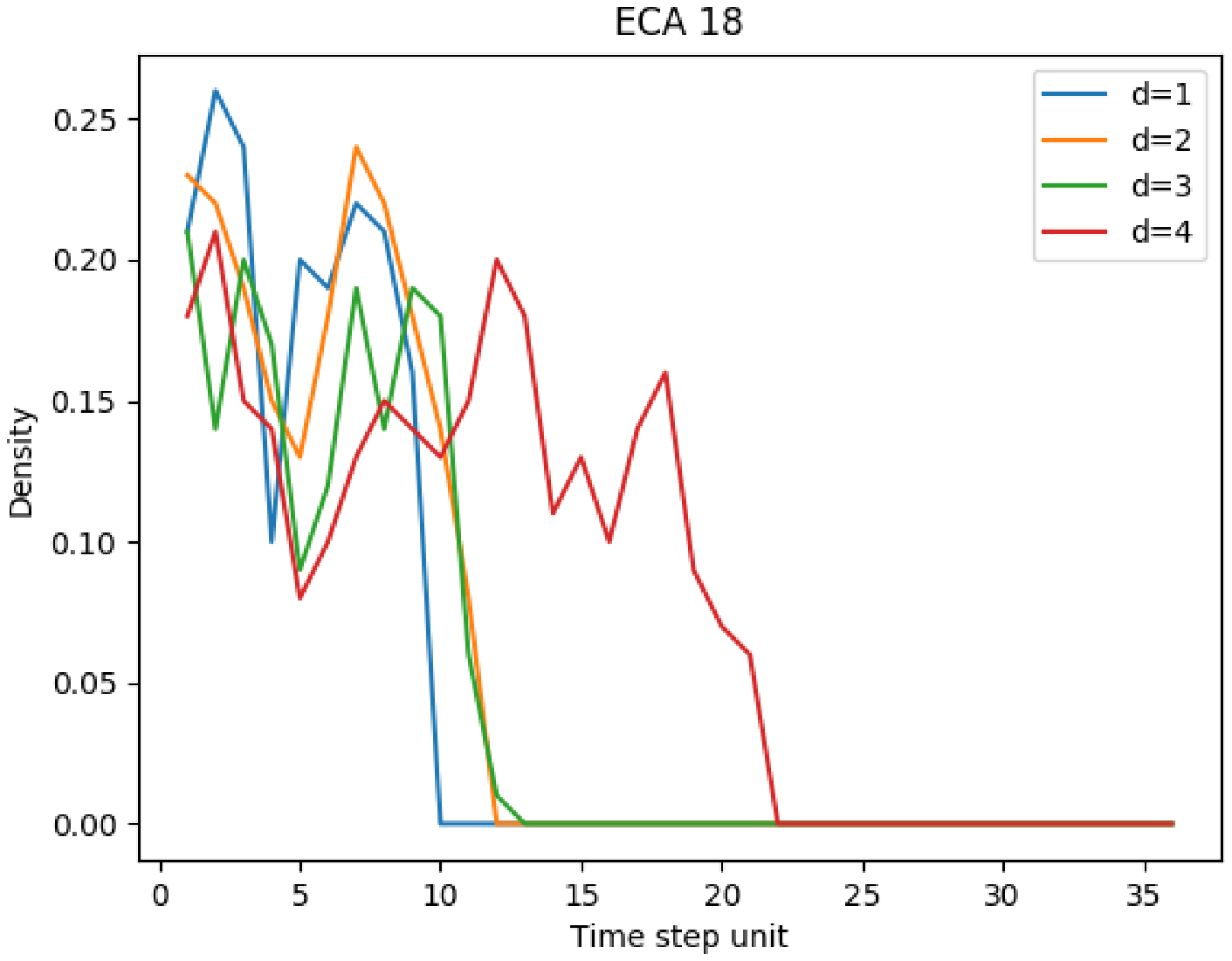} & \includegraphics[width=45mm]{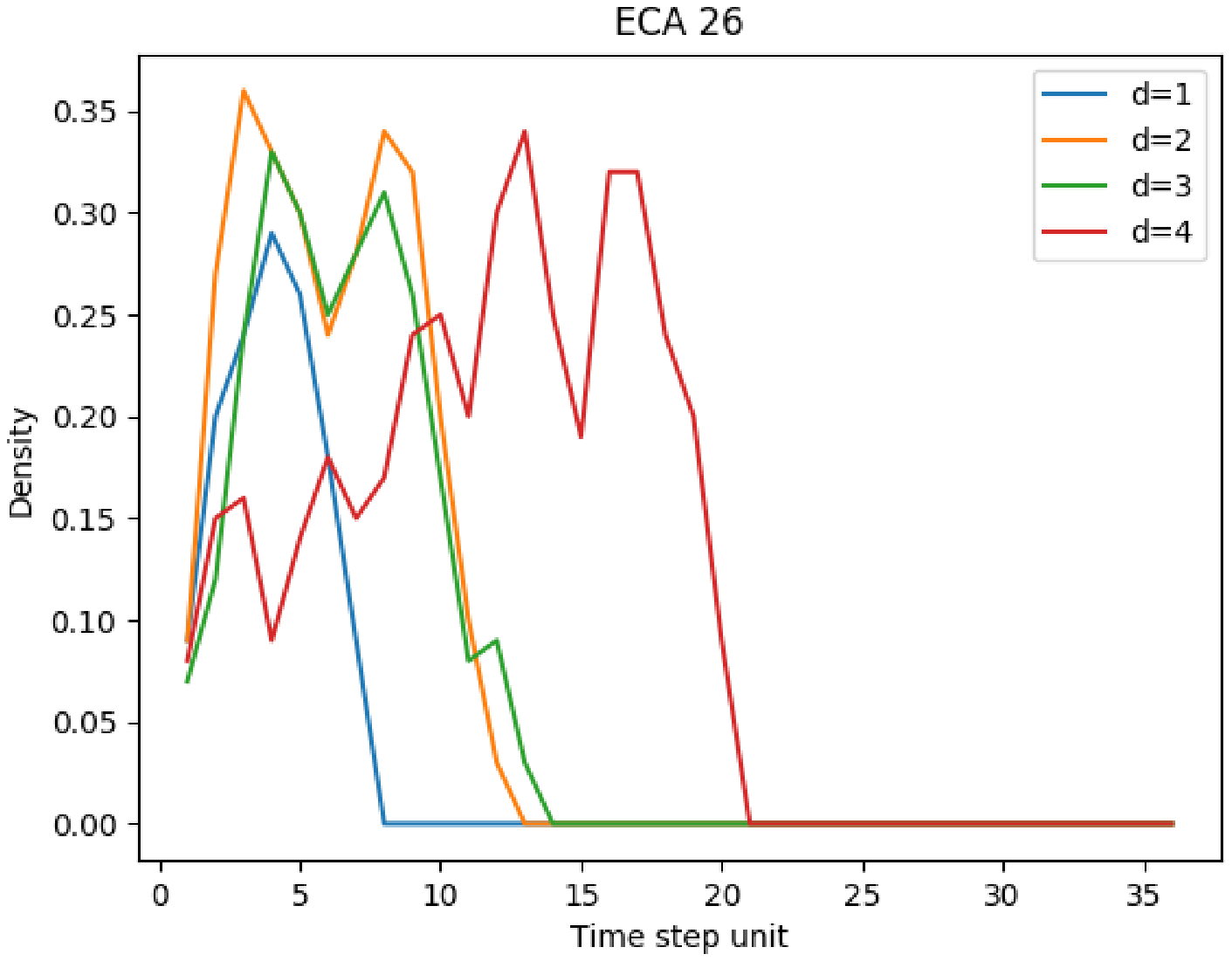} &
\includegraphics[width=45mm]{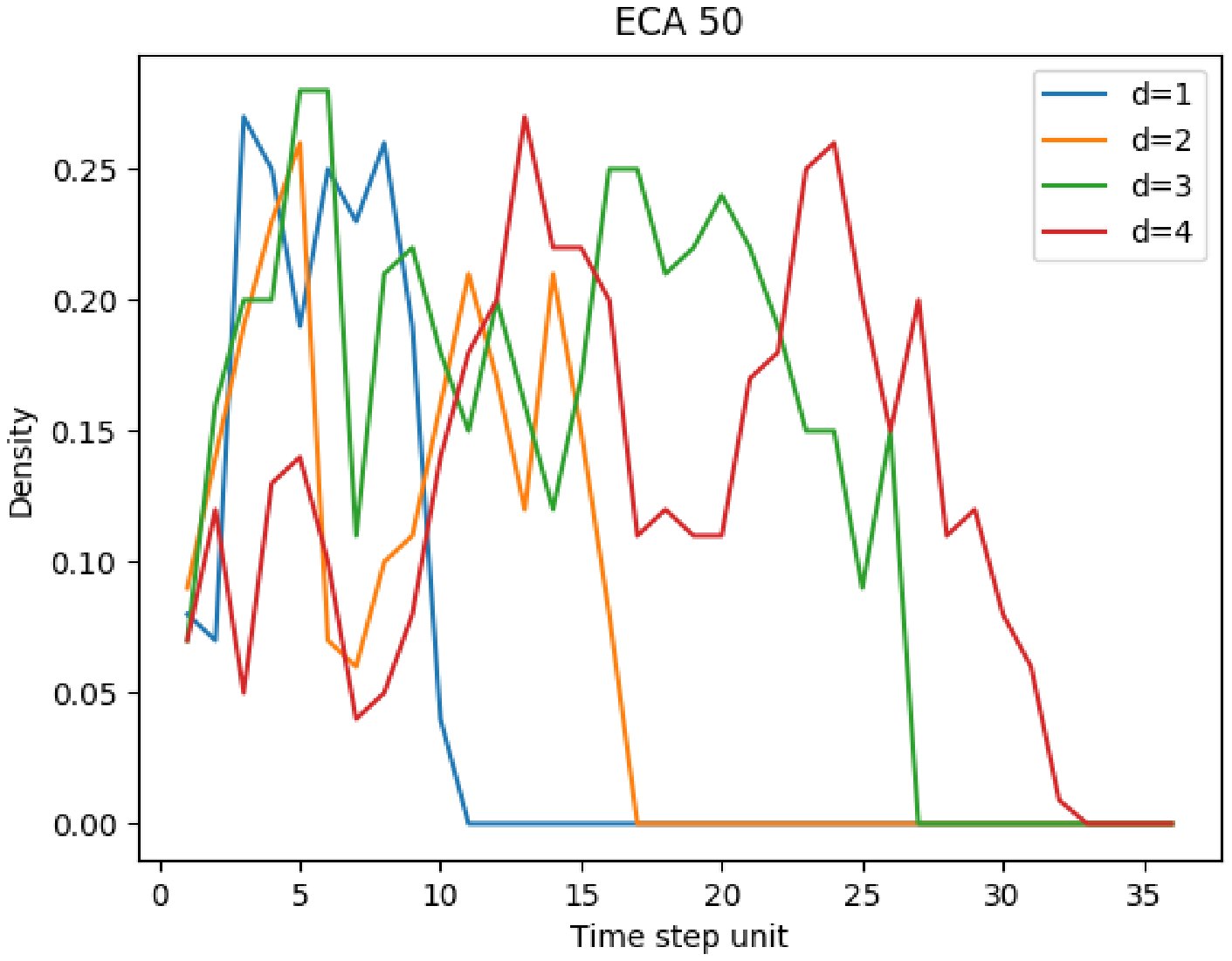} & \includegraphics[width=45mm]{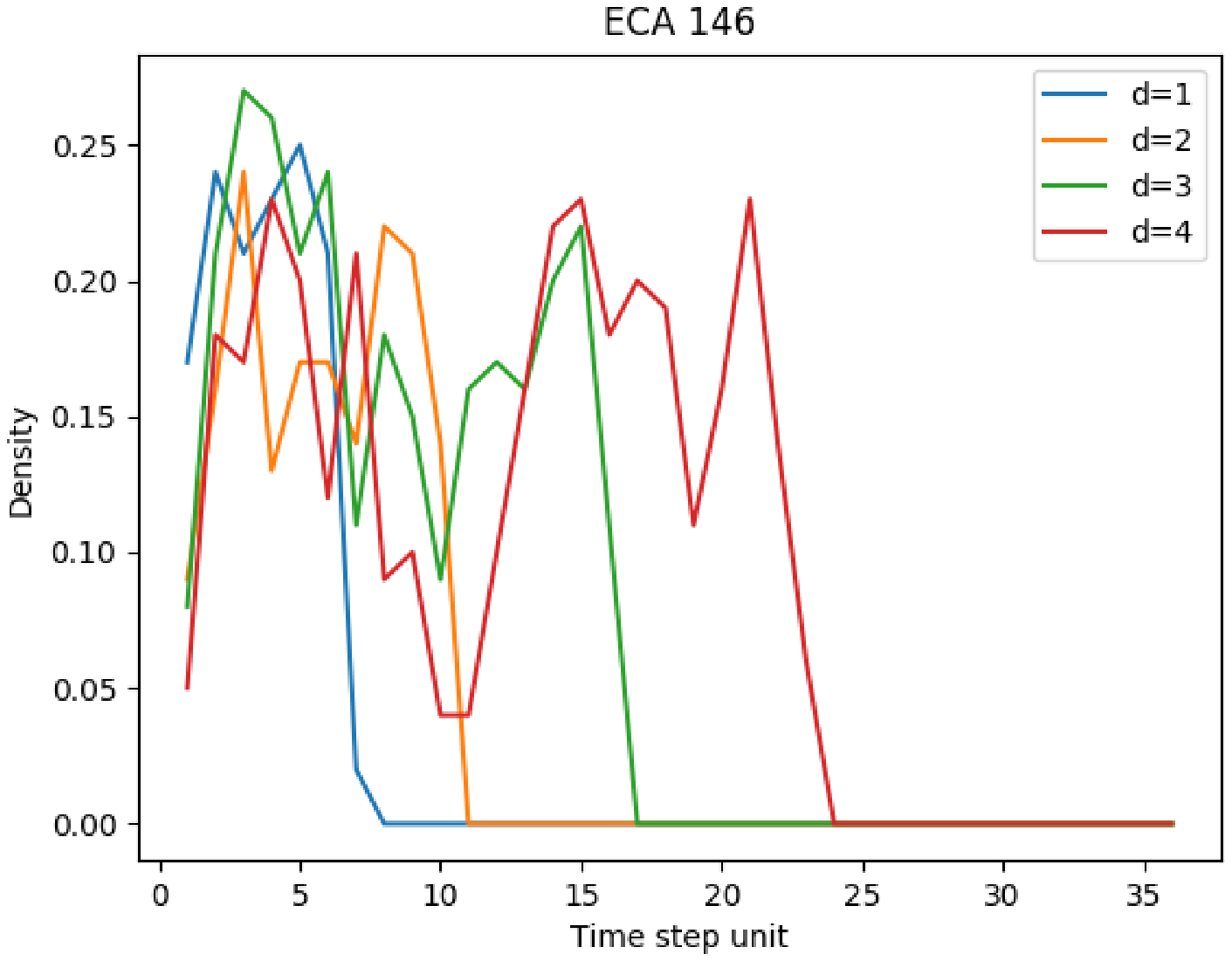} \\
\end{tabular}
\caption{The plot shows the profile of density parameter as a function of time steps with changing delay parameter for ECA rules. Here, the plot works with critical probabilistic loss of information perturbation rate $\iota^c$.}
\label{c2}
\end{figure*}

Now, this finding is mapped with the best candidate rule among ECA $18$,~$26$,~$50$,~$146$ for modelling the Baduria rioting dynamics. To compare the CA dynamics and Baduria riot dynamics, here, we let the system evolve starting from a single state `$1$' seed and average the density parameter value for every $100$ time steps which defines one single time unit ($\therefore$ $1$ time step $\approx$ 15 minute). Therefore, the {\em normalized density parameter} and {\em normalized overall intensity} of Baduria riot are the parameters for comparison in this study. Note that, normalized density parameter is also calculated following the similar procedure of calculating normalized overall intensity parameter. Fig.~\ref{c1} shows the profile of normalized density parameter or normalized overall intensity of Baduria riot as a function of the time series. According to Fig.~\ref{c1}, ECAs $26$ and $146$ show similar dynamics with Baduria riot, however, ECAs $18$ and $50$ come up with `late' convergence dynamics for critical probabilistic loss of information perturbation rate ($\iota^c$) where $d = 1$. We identified the best fitting CA model using the standard {\em root mean square deviation} = $\sqrt{\dfrac{\sum_{t=1}^{T} (x_{1,t}-x_{2,t})^2}{T}}$ where $x_{1,t}$ and $x_{2,t}$ depict the time series normalized overall intensity of Baduria riots and density parameter value of CA model respectively. Here, ECA $26$ ($d$ = $1$ and $\iota^c$ = $.51$) shows the best fitting attitude with Baduria riot data where the root mean square deviation = $0.064$. Note that, ECA $18$,~$50$ and $146$ are associated with root mean square derivation value $0.129$,~$0.138$ and $0.071$ respectively. The best fitting CA with $d$ = $1$ depicts the evidence of no major presence of organized communal force in the rioting society. Note that, the field data also reflects this absence. Fig.~\ref{c2} depicts the evidence associated with `late' convergence with increasing value of delay perturbation parameter for these ECA rules. Without presence of organized communal force, the rioting society reacts spontaneously and a simple up and down dynamics is reflected. However, increasing rebound dynamics is observed for increasing value of delay perturbation parameter, see Fig.~\ref{c2} as evidence. As an insight, the rebound dynamics indicates that organized communal force plays role for regenerating the spontaneity of rioting in the society. 

The proposed model only verifies rioting dynamics of West Bengal event. However, as a discussion, ECA $146$ depicts similar sudden-down dynamics of $2011$ London riots \cite{Davies2013211}, for evidence see Fig.~\ref{c2}. Moreover, Berestycki et. al \cite{Berestycki11} have analysed sudden spike dynamics to represent strong exogenous factor and slower but steady increase to reflect endogenous factors in rioting dynamics. In this context, ECA $18$ and $50$ shows similar sudden spike and steady growth dynamics respectively, see Fig.~\ref{c2} for evidence. However, proper understanding about this similar signature behaviour of the proposed CA system and exogenous-endogenous factors is still an open question for us. Now, depending on the wide variety of results, the study strongly indicates that this model is relevant for other rioting dynamics.

To sumup, the study reflects the modelling aspects of the internal convergence rioting dynamics with respect to the sociological factors - presence of anti-riot population as well as organizational presence of communal forces. One may argue about the presence of other sociological factors in the rioting dynamics, however, our target is to propose a simple model which can able to capture the complex rioting dynamics. To validate our argument, we quote from  Burbeck et al. \cite{9989878} which is the pioneering work on epidemiological mathematical riot modelling.

\begin{quote}

\textit{``First efforts at model building in a new field necessarily encounter the risk of oversimplification, yet if the models are not kept as simple as is practical they tend to become immune to falsification.\cite{9989878}"}

\end{quote}
Moreover, this simple local interaction model can be able to capture non-locality using delay perturbation. Here, the proposed CA system introduces non-uniformity with respect to information sharing aspects, however, the system is associated with uniform rule. Several non-uniform internal dynamics are also reflected in Baduria incident, i.e. here, three types of internal dynamics (first: looting incident
\footnote{On 3rd and 4th day of July 2017 there was attack and loot in the Hindu shops of the area, along with a small pan and cigarette shop nearby. He claimed the loot was of 22lakhs during the incident. Out of sorrow he expressed that the small pan and cigarette shop received its reimbursement whereas others including him did not receive. In the Trimohini area the Hindu and Muslim shops are situated on opposite sides of the main road, due to which it is easy to identify the shops by their religion. In the evening at around 8pm of 3rd July there was an attack in the Rath Jatra at Trimohini which was followed by the loot at mid-night. Following this incident there were attacks and loots in the Muslim shops of the area on 5th July. Name: Sambhu Seth; Address: Trimohini, Basirhat; Occupation: Owner of Private Eye Hospital; Interview date, time: 20th October, 2019, 12:00pm}; 
second: local resident vs refugees 
\footnote{He said that the Hindu and Muslim residents of the area were united and wanted peace. He further accused towards the Bangladeshi Hindu refugees of the area. He also added that these Bangladeshi Hindu refugees have no permanent residence or land and mostly inhabits in the bustees of adjoining railway land. Along with these Bangladeshi Hindu refugees may be outsiders were involved in the riot. Name: Jallaluddin; Address: Bhabla halt, Basirhat; Occupation: Cycle garage owner; Interview date, time: 20th October, 2019, 3:00pm}; 
third: local people vs police 
\footnote{He said that there was no such Hindu-Muslim clash incident in Baduria not like Bashirhat. Mainly there was clash between police and the local people. Name: Muslim-Shopkeeper; Address: Baduria Chowmata (near Baduria police station); Interview date, time:  20th October, 2019, 6:00pm}) 
are reflected from the field data. The current construction of CA model with uniform rule is unable to address this microscopic dynamics. Therefore, this research can be extended to explore the dynamics of the CA model with non-uniform rule. In terms of dynamical systems, the model adopts time varying system rather than linear time invariant system in the journey towards full fledged non-linear dynamics.

\textbf{Acknowledgements} This research is partially supported by Impactful Policy Research in Social Science (IMPRESS) under Indian Council of Social Science Research, Govt. of India (P3271/220).

The authors are grateful to Priyadarsini Sinha for her work as a field data collecting team member. The authors are also grateful to Prof. Subhasis Bandopadhyay for enlightening discussions on this topic.

\bibliography{ref}

\begin{thebibliography}{66}%
\makeatletter
\providecommand \@ifxundefined [1]{%
 \@ifx{#1\undefined}
}%
\providecommand \@ifnum [1]{%
 \ifnum #1\expandafter \@firstoftwo
 \else \expandafter \@secondoftwo
 \fi
}%
\providecommand \@ifx [1]{%
 \ifx #1\expandafter \@firstoftwo
 \else \expandafter \@secondoftwo
 \fi
}%
\providecommand \natexlab [1]{#1}%
\providecommand \enquote  [1]{``#1''}%
\providecommand \bibnamefont  [1]{#1}%
\providecommand \bibfnamefont [1]{#1}%
\providecommand \citenamefont [1]{#1}%
\providecommand \href@noop [0]{\@secondoftwo}%
\providecommand \href [0]{\begingroup \@sanitize@url \@href}%
\providecommand \@href[1]{\@@startlink{#1}\@@href}%
\providecommand \@@href[1]{\endgroup#1\@@endlink}%
\providecommand \@sanitize@url [0]{\catcode `\\12\catcode `\$12\catcode
  `\&12\catcode `\#12\catcode `\^12\catcode `\_12\catcode `\%12\relax}%
\providecommand \@@startlink[1]{}%
\providecommand \@@endlink[0]{}%
\providecommand \url  [0]{\begingroup\@sanitize@url \@url }%
\providecommand \@url [1]{\endgroup\@href {#1}{\urlprefix }}%
\providecommand \urlprefix  [0]{URL }%
\providecommand \Eprint [0]{\href }%
\providecommand \doibase [0]{http://dx.doi.org/}%
\providecommand \selectlanguage [0]{\@gobble}%
\providecommand \bibinfo  [0]{\@secondoftwo}%
\providecommand \bibfield  [0]{\@secondoftwo}%
\providecommand \translation [1]{[#1]}%
\providecommand \BibitemOpen [0]{}%
\providecommand \bibitemStop [0]{}%
\providecommand \bibitemNoStop [0]{.\EOS\space}%
\providecommand \EOS [0]{\spacefactor3000\relax}%
\providecommand \BibitemShut  [1]{\csname bibitem#1\endcsname}%
\let\auto@bib@innerbib\@empty
\bibitem [{\citenamefont {Midlarsky}(1978)}]{midlarsky1978}%
  \BibitemOpen
  \bibfield  {author} {\bibinfo {author} {\bibfnamefont {M.~I.}\ \bibnamefont
  {Midlarsky}},\ }\href@noop {} {\bibfield  {journal} {\bibinfo  {journal}
  {American Political Science Review}\ }\textbf {\bibinfo {volume} {72}},\
  \bibinfo {pages} {996} (\bibinfo {year} {1978})}\BibitemShut {NoStop}%
\bibitem [{\citenamefont {Govea}\ and\ \citenamefont {West}(1981)}]{Rodger26}%
  \BibitemOpen
  \bibfield  {author} {\bibinfo {author} {\bibfnamefont {R.~M.}\ \bibnamefont
  {Govea}}\ and\ \bibinfo {author} {\bibfnamefont {G.~T.}\ \bibnamefont
  {West}},\ }\href@noop {} {\bibfield  {journal} {\bibinfo  {journal} {The
  Journal of Conflict Resolution}\ }\textbf {\bibinfo {volume} {25}},\ \bibinfo
  {pages} {349} (\bibinfo {year} {1981})}\BibitemShut {NoStop}%
\bibitem [{\citenamefont {Bohstedt}\ and\ \citenamefont
  {Williams}(1988)}]{Bohstedt21}%
  \BibitemOpen
  \bibfield  {author} {\bibinfo {author} {\bibfnamefont {J.}~\bibnamefont
  {Bohstedt}}\ and\ \bibinfo {author} {\bibfnamefont {D.~E.}\ \bibnamefont
  {Williams}},\ }\href@noop {} {\bibfield  {journal} {\bibinfo  {journal} {The
  Journal of Interdisciplinary History}\ }\textbf {\bibinfo {volume} {19}},\
  \bibinfo {pages} {1} (\bibinfo {year} {1988})}\BibitemShut {NoStop}%
\bibitem [{\citenamefont {Charlesworth}(1994)}]{charlesworth1994}%
  \BibitemOpen
  \bibfield  {author} {\bibinfo {author} {\bibfnamefont {A.}~\bibnamefont
  {Charlesworth}},\ }\href@noop {} {\bibfield  {journal} {\bibinfo  {journal}
  {Rural History}\ }\textbf {\bibinfo {volume} {5}},\ \bibinfo {pages} {1–22}
  (\bibinfo {year} {1994})}\BibitemShut {NoStop}%
\bibitem [{\citenamefont {Myers}(2010)}]{Daniel4v}%
  \BibitemOpen
  \bibfield  {author} {\bibinfo {author} {\bibfnamefont {D.}~\bibnamefont
  {Myers}},\ }\href@noop {} {\bibfield  {journal} {\bibinfo  {journal}
  {Mobilization: An International Quarterly}\ }\textbf {\bibinfo {volume}
  {15}},\ \bibinfo {pages} {289} (\bibinfo {year} {2010})}\BibitemShut
  {NoStop}%
\bibitem [{\citenamefont {Myers}(1997)}]{Myers57454}%
  \BibitemOpen
  \bibfield  {author} {\bibinfo {author} {\bibfnamefont {D.~J.}\ \bibnamefont
  {Myers}},\ }\href@noop {} {\bibfield  {journal} {\bibinfo  {journal}
  {American Sociological Review}\ }\textbf {\bibinfo {volume} {62}},\ \bibinfo
  {pages} {94} (\bibinfo {year} {1997})}\BibitemShut {NoStop}%
\bibitem [{\citenamefont {Das}(1990)}]{Das7477}%
  \BibitemOpen
  \bibfield  {author} {\bibinfo {author} {\bibfnamefont {S.}~\bibnamefont
  {Das}},\ }\href@noop {} {\bibfield  {journal} {\bibinfo  {journal} {Social
  Scientist}\ }\textbf {\bibinfo {volume} {18}},\ \bibinfo {pages} {21}
  (\bibinfo {year} {1990})}\BibitemShut {NoStop}%
\bibitem [{\citenamefont {Das}(1993)}]{das1993communal}%
  \BibitemOpen
  \bibfield  {author} {\bibinfo {author} {\bibfnamefont {S.}~\bibnamefont
  {Das}},\ }\href@noop {} {\emph {\bibinfo {title} {Communal Riots in Bengal,
  1905-1947}}}\ (\bibinfo  {publisher} {Oxford University Press},\ \bibinfo
  {year} {1993})\BibitemShut {NoStop}%
\bibitem [{\citenamefont {Das}(2000)}]{article1}%
  \BibitemOpen
  \bibfield  {author} {\bibinfo {author} {\bibfnamefont {S.}~\bibnamefont
  {Das}},\ }\href {\doibase 10.1017/S0026749X0000336X} {\bibfield  {journal}
  {\bibinfo  {journal} {Modern Asian Studies}\ }\textbf {\bibinfo {volume}
  {34}},\ \bibinfo {pages} {281 } (\bibinfo {year} {2000})}\BibitemShut
  {NoStop}%
\bibitem [{\citenamefont {Burbeck}\ \emph {et~al.}(1978)\citenamefont
  {Burbeck}, \citenamefont {Raine},\ and\ \citenamefont {Stark}}]{9989878}%
  \BibitemOpen
  \bibfield  {author} {\bibinfo {author} {\bibfnamefont {S.~L.}\ \bibnamefont
  {Burbeck}}, \bibinfo {author} {\bibfnamefont {W.~J.}\ \bibnamefont {Raine}},
  \ and\ \bibinfo {author} {\bibfnamefont {M.~J.~A.}\ \bibnamefont {Stark}},\
  }\href@noop {} {\bibfield  {journal} {\bibinfo  {journal} {The Journal of
  Mathematical Sociology}\ }\textbf {\bibinfo {volume} {6}},\ \bibinfo {pages}
  {1} (\bibinfo {year} {1978})}\BibitemShut {NoStop}%
\bibitem [{\citenamefont {Bonnasse-Gahot}\ \emph {et~al.}(2018)\citenamefont
  {Bonnasse-Gahot}, \citenamefont {Berestycki}, \citenamefont {Depuiset},
  \citenamefont {Gordon}, \citenamefont {Roché}, \citenamefont {Rodriguez},\
  and\ \citenamefont {Nadal}}]{Laurent211}%
  \BibitemOpen
  \bibfield  {author} {\bibinfo {author} {\bibfnamefont {L.}~\bibnamefont
  {Bonnasse-Gahot}}, \bibinfo {author} {\bibfnamefont {H.}~\bibnamefont
  {Berestycki}}, \bibinfo {author} {\bibfnamefont {M.-A.}\ \bibnamefont
  {Depuiset}}, \bibinfo {author} {\bibfnamefont {M.~B.}\ \bibnamefont
  {Gordon}}, \bibinfo {author} {\bibfnamefont {S.}~\bibnamefont {Roché}},
  \bibinfo {author} {\bibfnamefont {N.}~\bibnamefont {Rodriguez}}, \ and\
  \bibinfo {author} {\bibfnamefont {J.-P.}\ \bibnamefont {Nadal}},\ }\href@noop
  {} {\bibfield  {journal} {\bibinfo  {journal} {Scientific Reports}\ }\textbf
  {\bibinfo {volume} {8}} (\bibinfo {year} {2018})}\BibitemShut {NoStop}%
\bibitem [{\citenamefont {Davies}\ \emph {et~al.}(2013)\citenamefont {Davies},
  \citenamefont {Fry}, \citenamefont {Wilson},\ and\ \citenamefont
  {Bishop}}]{Davies2013211}%
  \BibitemOpen
  \bibfield  {author} {\bibinfo {author} {\bibfnamefont {T.~P.}\ \bibnamefont
  {Davies}}, \bibinfo {author} {\bibfnamefont {H.~M.}\ \bibnamefont {Fry}},
  \bibinfo {author} {\bibfnamefont {A.~G.}\ \bibnamefont {Wilson}}, \ and\
  \bibinfo {author} {\bibfnamefont {S.~R.}\ \bibnamefont {Bishop}},\
  }\href@noop {} {\bibfield  {journal} {\bibinfo  {journal} {Scientific
  Reports}\ }\textbf {\bibinfo {volume} {3}} (\bibinfo {year}
  {2013})}\BibitemShut {NoStop}%
\bibitem [{\citenamefont {Berestycki}\ \emph {et~al.}(2015)\citenamefont
  {Berestycki}, \citenamefont {Nadal},\ and\ \citenamefont
  {Rodiguez}}]{Berestycki11}%
  \BibitemOpen
  \bibfield  {author} {\bibinfo {author} {\bibfnamefont {H.}~\bibnamefont
  {Berestycki}}, \bibinfo {author} {\bibfnamefont {J.-P.}\ \bibnamefont
  {Nadal}}, \ and\ \bibinfo {author} {\bibfnamefont {N.}~\bibnamefont
  {Rodiguez}},\ }\href@noop {} {\bibfield  {journal} {\bibinfo  {journal}
  {Networks \& Heterogeneous Media}\ }\textbf {\bibinfo {volume} {10}},\
  \bibinfo {pages} {443} (\bibinfo {year} {2015})}\BibitemShut {NoStop}%
\bibitem [{\citenamefont {Pitcher}\ \emph {et~al.}(1978)\citenamefont
  {Pitcher}, \citenamefont {Hamblin},\ and\ \citenamefont {Miller}}]{Brian9}%
  \BibitemOpen
  \bibfield  {author} {\bibinfo {author} {\bibfnamefont {B.~L.}\ \bibnamefont
  {Pitcher}}, \bibinfo {author} {\bibfnamefont {R.~L.}\ \bibnamefont
  {Hamblin}}, \ and\ \bibinfo {author} {\bibfnamefont {J.~L.~L.}\ \bibnamefont
  {Miller}},\ }\href@noop {} {\bibfield  {journal} {\bibinfo  {journal}
  {American Sociological Review}\ }\textbf {\bibinfo {volume} {43}},\ \bibinfo
  {pages} {23} (\bibinfo {year} {1978})}\BibitemShut {NoStop}%
\bibitem [{\citenamefont {Myers}(2000)}]{Daniel10}%
  \BibitemOpen
  \bibfield  {author} {\bibinfo {author} {\bibfnamefont {D.}~\bibnamefont
  {Myers}},\ }\href@noop {} {\bibfield  {journal} {\bibinfo  {journal}
  {American Journal of Sociology}\ }\textbf {\bibinfo {volume} {106}},\
  \bibinfo {pages} {173} (\bibinfo {year} {2000})}\BibitemShut {NoStop}%
\bibitem [{\citenamefont {Braha}(2012)}]{Braha8}%
  \BibitemOpen
  \bibfield  {author} {\bibinfo {author} {\bibfnamefont {D.}~\bibnamefont
  {Braha}},\ }\href@noop {} {\bibfield  {journal} {\bibinfo  {journal} {PLoS
  ONE}\ }\textbf {\bibinfo {volume} {7}} (\bibinfo {year} {2012})}\BibitemShut
  {NoStop}%
\bibitem [{\citenamefont {Baudains}\ \emph {et~al.}(2013)\citenamefont
  {Baudains}, \citenamefont {Johnson},\ and\ \citenamefont
  {Braithwaite}}]{BAUDAINS2013211}%
  \BibitemOpen
  \bibfield  {author} {\bibinfo {author} {\bibfnamefont {P.}~\bibnamefont
  {Baudains}}, \bibinfo {author} {\bibfnamefont {S.~D.}\ \bibnamefont
  {Johnson}}, \ and\ \bibinfo {author} {\bibfnamefont {A.~M.}\ \bibnamefont
  {Braithwaite}},\ }\href@noop {} {\bibfield  {journal} {\bibinfo  {journal}
  {Applied Geography}\ }\textbf {\bibinfo {volume} {45}},\ \bibinfo {pages}
  {211 } (\bibinfo {year} {2013})}\BibitemShut {NoStop}%
\bibitem [{\citenamefont {Granovetter}(1978)}]{2778111}%
  \BibitemOpen
  \bibfield  {author} {\bibinfo {author} {\bibfnamefont {M.}~\bibnamefont
  {Granovetter}},\ }\href@noop {} {\bibfield  {journal} {\bibinfo  {journal}
  {American Journal of Sociology}\ }\textbf {\bibinfo {volume} {83}},\ \bibinfo
  {pages} {1420} (\bibinfo {year} {1978})}\BibitemShut {NoStop}%
\bibitem [{\citenamefont {Stark}\ \emph {et~al.}(1974)\citenamefont {Stark},
  \citenamefont {Raine}, \citenamefont {Burbeck},\ and\ \citenamefont
  {Davison}}]{10.2307/2094159}%
  \BibitemOpen
  \bibfield  {author} {\bibinfo {author} {\bibfnamefont {M.~J.~A.}\
  \bibnamefont {Stark}}, \bibinfo {author} {\bibfnamefont {W.~J.}\ \bibnamefont
  {Raine}}, \bibinfo {author} {\bibfnamefont {S.~L.}\ \bibnamefont {Burbeck}},
  \ and\ \bibinfo {author} {\bibfnamefont {K.~K.}\ \bibnamefont {Davison}},\
  }\href@noop {} {\bibfield  {journal} {\bibinfo  {journal} {American
  Sociological Review}\ }\textbf {\bibinfo {volume} {39}},\ \bibinfo {pages}
  {865} (\bibinfo {year} {1974})}\BibitemShut {NoStop}%
\bibitem [{\citenamefont {González-Bailón}\ \emph {et~al.}(2011)\citenamefont
  {González-Bailón}, \citenamefont {Borge-Holthoefer}, \citenamefont
  {Rivero},\ and\ \citenamefont {Moreno}}]{Sandra11}%
  \BibitemOpen
  \bibfield  {author} {\bibinfo {author} {\bibfnamefont {S.}~\bibnamefont
  {González-Bailón}}, \bibinfo {author} {\bibfnamefont {J.}~\bibnamefont
  {Borge-Holthoefer}}, \bibinfo {author} {\bibfnamefont {A.}~\bibnamefont
  {Rivero}}, \ and\ \bibinfo {author} {\bibfnamefont {Y.}~\bibnamefont
  {Moreno}},\ }\href@noop {} {\bibfield  {journal} {\bibinfo  {journal}
  {Scientific Reports}\ }\textbf {\bibinfo {volume} {1}} (\bibinfo {year}
  {2011})}\BibitemShut {NoStop}%
\bibitem [{\citenamefont {Wolfram}(1986)}]{wolfram86}%
  \BibitemOpen
  \bibfield  {author} {\bibinfo {author} {\bibfnamefont {S.}~\bibnamefont
  {Wolfram}},\ }\href@noop {} {\emph {\bibinfo {title} {Theory and applications
  of cellular automata}}}\ (\bibinfo  {publisher} {World Scientific},\ \bibinfo
  {address} {Singapore, ISBN 9971-50-124-4 pbk},\ \bibinfo {year}
  {1986})\BibitemShut {NoStop}%
\bibitem [{\citenamefont {Li}\ and\ \citenamefont
  {Packard}(1990)}]{Li90thestructure}%
  \BibitemOpen
  \bibfield  {author} {\bibinfo {author} {\bibfnamefont {W.}~\bibnamefont
  {Li}}\ and\ \bibinfo {author} {\bibfnamefont {N.}~\bibnamefont {Packard}},\
  }\href@noop {} {\bibfield  {journal} {\bibinfo  {journal} {Complex Systems}\
  }\textbf {\bibinfo {volume} {4}},\ \bibinfo {pages} {281} (\bibinfo {year}
  {1990})}\BibitemShut {NoStop}%
\bibitem [{\citenamefont {Fat{\`{e}}s}(2014)}]{jcaFates14}%
  \BibitemOpen
  \bibfield  {author} {\bibinfo {author} {\bibfnamefont {N.}~\bibnamefont
  {Fat{\`{e}}s}},\ }\href@noop {} {\bibfield  {journal} {\bibinfo  {journal}
  {J. Cellular Automata}\ }\textbf {\bibinfo {volume} {9}},\ \bibinfo {pages}
  {387} (\bibinfo {year} {2014})}\BibitemShut {NoStop}%
\bibitem [{\citenamefont {Roy}(2019)}]{ROY2019600}%
  \BibitemOpen
  \bibfield  {author} {\bibinfo {author} {\bibfnamefont {S.}~\bibnamefont
  {Roy}},\ }\href@noop {} {\bibfield  {journal} {\bibinfo  {journal} {Physica
  A: Statistical Mechanics and its Applications}\ }\textbf {\bibinfo {volume}
  {515}},\ \bibinfo {pages} {600 } (\bibinfo {year} {2019})}\BibitemShut
  {NoStop}%
\bibitem [{\citenamefont {Sethi}\ \emph {et~al.}(2016)\citenamefont {Sethi},
  \citenamefont {Roy},\ and\ \citenamefont {Das}}]{SethiRD16}%
  \BibitemOpen
  \bibfield  {author} {\bibinfo {author} {\bibfnamefont {B.}~\bibnamefont
  {Sethi}}, \bibinfo {author} {\bibfnamefont {S.}~\bibnamefont {Roy}}, \ and\
  \bibinfo {author} {\bibfnamefont {S.}~\bibnamefont {Das}},\ }\href@noop {}
  {\bibfield  {journal} {\bibinfo  {journal} {Complexity}\ }\textbf {\bibinfo
  {volume} {21}},\ \bibinfo {pages} {370} (\bibinfo {year} {2016})}\BibitemShut
  {NoStop}%
\bibitem [{\citenamefont {Schönfisch}\ and\ \citenamefont
  {de~Roos}(1999)}]{SCHONFISCH1999123}%
  \BibitemOpen
  \bibfield  {author} {\bibinfo {author} {\bibfnamefont {B.}~\bibnamefont
  {Schönfisch}}\ and\ \bibinfo {author} {\bibfnamefont {A.}~\bibnamefont
  {de~Roos}},\ }\href@noop {} {\bibfield  {journal} {\bibinfo  {journal}
  {Biosystems}\ }\textbf {\bibinfo {volume} {51}},\ \bibinfo {pages} {123 }
  (\bibinfo {year} {1999})}\BibitemShut {NoStop}%
\bibitem [{\citenamefont {Blok}\ and\ \citenamefont
  {Bergersen}(1999)}]{PhysRevE5976}%
  \BibitemOpen
  \bibfield  {author} {\bibinfo {author} {\bibfnamefont {H.~J.}\ \bibnamefont
  {Blok}}\ and\ \bibinfo {author} {\bibfnamefont {B.}~\bibnamefont
  {Bergersen}},\ }\href@noop {} {\bibfield  {journal} {\bibinfo  {journal}
  {Phys. Rev. E}\ }\textbf {\bibinfo {volume} {59}},\ \bibinfo {pages} {3876}
  (\bibinfo {year} {1999})}\BibitemShut {NoStop}%
\bibitem [{\citenamefont {Reia}\ and\ \citenamefont
  {Kinouchi}(2015)}]{PhysRevE910421}%
  \BibitemOpen
  \bibfield  {author} {\bibinfo {author} {\bibfnamefont {S.~M.}\ \bibnamefont
  {Reia}}\ and\ \bibinfo {author} {\bibfnamefont {O.}~\bibnamefont
  {Kinouchi}},\ }\href@noop {} {\bibfield  {journal} {\bibinfo  {journal}
  {Phys. Rev. E}\ }\textbf {\bibinfo {volume} {91}},\ \bibinfo {pages} {042110}
  (\bibinfo {year} {2015})}\BibitemShut {NoStop}%
\bibitem [{\citenamefont {Bour{\'e}}\ \emph {et~al.}(2012)\citenamefont
  {Bour{\'e}}, \citenamefont {Fat{\`e}s},\ and\ \citenamefont
  {Chevrier}}]{Bou2012}%
  \BibitemOpen
  \bibfield  {author} {\bibinfo {author} {\bibfnamefont {O.}~\bibnamefont
  {Bour{\'e}}}, \bibinfo {author} {\bibfnamefont {N.}~\bibnamefont
  {Fat{\`e}s}}, \ and\ \bibinfo {author} {\bibfnamefont {V.}~\bibnamefont
  {Chevrier}},\ }\href@noop {} {\bibfield  {journal} {\bibinfo  {journal}
  {Natural Computing}\ }\textbf {\bibinfo {volume} {11}},\ \bibinfo {pages}
  {553} (\bibinfo {year} {2012})}\BibitemShut {NoStop}%
\bibitem [{\citenamefont {Sethi}\ \emph {et~al.}(2018)\citenamefont {Sethi},
  \citenamefont {Roy},\ and\ \citenamefont {Das}}]{130519}%
  \BibitemOpen
  \bibfield  {author} {\bibinfo {author} {\bibfnamefont {B.}~\bibnamefont
  {Sethi}}, \bibinfo {author} {\bibfnamefont {S.}~\bibnamefont {Roy}}, \ and\
  \bibinfo {author} {\bibfnamefont {S.}~\bibnamefont {Das}},\ }\href@noop {}
  {\bibfield  {journal} {\bibinfo  {journal} {Journal of Cellular Automata}\
  }\textbf {\bibinfo {volume} {13}},\ \bibinfo {pages} {527 } (\bibinfo {year}
  {2018})}\BibitemShut {NoStop}%
\bibitem [{\citenamefont {Redeker}\ \emph {et~al.}(2013)\citenamefont
  {Redeker}, \citenamefont {Adamatzky},\ and\ \citenamefont
  {Martinez}}]{MARKUS101}%
  \BibitemOpen
  \bibfield  {author} {\bibinfo {author} {\bibfnamefont {M.}~\bibnamefont
  {Redeker}}, \bibinfo {author} {\bibfnamefont {A.}~\bibnamefont {Adamatzky}},
  \ and\ \bibinfo {author} {\bibfnamefont {G.~J.}\ \bibnamefont {Martinez}},\
  }\href@noop {} {\bibfield  {journal} {\bibinfo  {journal} {International
  Journal of Modern Physics C}\ }\textbf {\bibinfo {volume} {24}},\ \bibinfo
  {pages} {1350010} (\bibinfo {year} {2013})}\BibitemShut {NoStop}%
\bibitem [{\citenamefont {Gravner}\ and\ \citenamefont
  {Griffeath}(2011)}]{Gravner2011}%
  \BibitemOpen
  \bibfield  {author} {\bibinfo {author} {\bibfnamefont {J.}~\bibnamefont
  {Gravner}}\ and\ \bibinfo {author} {\bibfnamefont {D.}~\bibnamefont
  {Griffeath}},\ }\href@noop {} {\bibfield  {journal} {\bibinfo  {journal}
  {Journal of Statistical Physics}\ }\textbf {\bibinfo {volume} {142}},\
  \bibinfo {pages} {168} (\bibinfo {year} {2011})}\BibitemShut {NoStop}%
\bibitem [{\citenamefont {Gravner}\ and\ \citenamefont
  {Griffeath}(2012)}]{GRAVNER201264}%
  \BibitemOpen
  \bibfield  {author} {\bibinfo {author} {\bibfnamefont {J.}~\bibnamefont
  {Gravner}}\ and\ \bibinfo {author} {\bibfnamefont {D.}~\bibnamefont
  {Griffeath}},\ }\href@noop {} {\bibfield  {journal} {\bibinfo  {journal}
  {Theoretical Computer Science}\ }\textbf {\bibinfo {volume} {466}},\ \bibinfo
  {pages} {64} (\bibinfo {year} {2012})}\BibitemShut {NoStop}%
\bibitem [{\citenamefont {Wolfram}(1984)}]{WOLFRAM19841}%
  \BibitemOpen
  \bibfield  {author} {\bibinfo {author} {\bibfnamefont {S.}~\bibnamefont
  {Wolfram}},\ }\href {\doibase https://doi.org/10.1016/0167-2789(84)90245-8}
  {\bibfield  {journal} {\bibinfo  {journal} {Physica D: Nonlinear Phenomena}\
  }\textbf {\bibinfo {volume} {10}},\ \bibinfo {pages} {1 } (\bibinfo {year}
  {1984})}\BibitemShut {NoStop}%
\bibitem [{\citenamefont {Lang}\ and\ \citenamefont
  {Sterck}(2014)}]{LANG201412}%
  \BibitemOpen
  \bibfield  {author} {\bibinfo {author} {\bibfnamefont {J.}~\bibnamefont
  {Lang}}\ and\ \bibinfo {author} {\bibfnamefont {H.~D.}\ \bibnamefont
  {Sterck}},\ }\href@noop {} {\bibfield  {journal} {\bibinfo  {journal}
  {Mathematical Social Sciences}\ }\textbf {\bibinfo {volume} {69}},\ \bibinfo
  {pages} {12 } (\bibinfo {year} {2014})}\BibitemShut {NoStop}%
\bibitem [{ind(2016)}]{indiatoday2016}%
  \BibitemOpen
  \href
  {https://www.indiatoday.in/india/story/west-bengal-dhulagarh-howrah-clashes-religious-procession-358686-2016-12-20}
  {\bibfield  {journal} {\bibinfo  {journal} {INDIA TODAY}\ } (\bibinfo {year}
  {2016})}\BibitemShut {NoStop}%
\bibitem [{hin(2018)}]{hindustantimes2018}%
  \BibitemOpen
  \href
  {https://www.hindustantimes.com/kolkata/asansol-imam-who-lost-his-son-during-riots-appeals-for-peace-and-restraint/story-VSdUXOVkBPtEInB9tdIizH.html}
  {\bibfield  {journal} {\bibinfo  {journal} {hindustan times}\ } (\bibinfo
  {year} {2018})}\BibitemShut {NoStop}%
\bibitem [{bbc(2017)}]{bbc2017}%
  \BibitemOpen
  \href {https://www.bbc.com/news/world-asia-india-40553993} {\bibfield
  {journal} {\bibinfo  {journal} {BBC India}\ } (\bibinfo {year}
  {2017})}\BibitemShut {NoStop}%
\bibitem [{fin(2017)}]{financialexpress2017}%
  \BibitemOpen
  \href
  {https://www.financialexpress.com/india-news/baduria-basirhat-riot-how-muslims-pooled-money-to-help-hindus-targetted-by-rioters-west-bengal-mamata-banerjee-trinamool-bjp/755604/}
  {\bibfield  {journal} {\bibinfo  {journal} {FINANCIAL EXPRESS}\ } (\bibinfo
  {year} {2017})}\BibitemShut {NoStop}%
\bibitem [{new(2017)}]{news182017}%
  \BibitemOpen
  \href
  {https://www.news18.com/news/india/after-clashes-disturbed-their-lives-hindu-muslim-join-hands-to-rebuild-basirhat-1456189.html}
  {\bibfield  {journal} {\bibinfo  {journal} {NEWS 18}\ } (\bibinfo {year}
  {2017})}\BibitemShut {NoStop}%
\bibitem [{EXP(2017)}]{EXPRESS17}%
  \BibitemOpen
  \href
  {http://cms.newindianexpress.com/nation/2017/jul/05/a-boys-blasphemy-triggers-communal-tinderbox-on-the-border-1624750.html}
  {\bibfield  {journal} {\bibinfo  {journal} {THE NEW INDIAN EXPRESS}\ }
  (\bibinfo {year} {2017})}\BibitemShut {NoStop}%
\bibitem [{NDT(2017)}]{NDTV}%
  \BibitemOpen
  \href
  {https://www.ndtv.com/india-news/bengals-basirhat-tense-after-police-lathicharge-several-injured-10-updates-1721517}
  {\bibfield  {journal} {\bibinfo  {journal} {NDTV}\ } (\bibinfo {year}
  {2017})}\BibitemShut {NoStop}%
\bibitem [{hin(2017{\natexlab{a}})}]{hina1}%
  \BibitemOpen
  \href
  {https://www.hindustantimes.com/kolkata/hindu-man-dies-in-west-bengal-violence-clashes-erupt-in-basirhat-again/story-O1j8VjMqPCdgSdCYQRyVVJ.html}
  {\bibfield  {journal} {\bibinfo  {journal} {hindustantimes}\ } (\bibinfo
  {year} {2017}{\natexlab{a}})}\BibitemShut {NoStop}%
\bibitem [{fir(2017)}]{firstpost}%
  \BibitemOpen
  \href
  {https://www.firstpost.com/india/west-bengal-communal-riots-situation-in-basirhat-under-control-and-normal-says-mamata-banerjee-govt-3840847.html}
  {\bibfield  {journal} {\bibinfo  {journal} {firstpost}\ } (\bibinfo {year}
  {2017})}\BibitemShut {NoStop}%
\bibitem [{hin(2017{\natexlab{b}})}]{hina2}%
  \BibitemOpen
  \href
  {https://indianexpress.com/article/india/fb-post-communal-violence-leave-west-bengal-town-divided-scarred-4737945/}
  {\bibfield  {journal} {\bibinfo  {journal} {Indianexpress}\ } (\bibinfo
  {year} {2017}{\natexlab{b}})}\BibitemShut {NoStop}%
\bibitem [{hin(2017{\natexlab{c}})}]{hina3}%
  \BibitemOpen
  \href
  {https://www.firstpost.com/india/basirhat-ground-report-residents-of-violence-torn-region-negotiate-new-communal-faultlines-3799785.html}
  {\bibfield  {journal} {\bibinfo  {journal} {Firstpost}\ } (\bibinfo {year}
  {2017}{\natexlab{c}})}\BibitemShut {NoStop}%
\bibitem [{hin(2017{\natexlab{d}})}]{hina4}%
  \BibitemOpen
  \href
  {https://www.hindustantimes.com/india-news/bengal-violence-basirhat-s-muslim-leaders-tried-to-pacify-the-rioting-mob-but-couldn-t/story-8O6XTqN3yhsvKlfYSDK2aK.html}
  {\bibfield  {journal} {\bibinfo  {journal} {Hindustantimes}\ } (\bibinfo
  {year} {2017}{\natexlab{d}})}\BibitemShut {NoStop}%
\bibitem [{hin(2017{\natexlab{e}})}]{hina5}%
  \BibitemOpen
  \href
  {https://www.asianage.com/metros/kolkata/050717/baduria-turns-into-battlefield-over-fb-post.html}
  {\bibfield  {journal} {\bibinfo  {journal} {The Asian Age}\ } (\bibinfo
  {year} {2017}{\natexlab{e}})}\BibitemShut {NoStop}%
\bibitem [{hin(2017{\natexlab{f}})}]{hina6}%
  \BibitemOpen
  \href
  {https://www.financialexpress.com/india-news/baduria-basirhat-west-bengal-riots-live-updates-communal-clashes-mamata-banerjee-trinamool-congress-bjp-alleges-2000-muslims-attacked-hindu-families/749764/}
  {\bibfield  {journal} {\bibinfo  {journal} {Financial Express}\ } (\bibinfo
  {year} {2017}{\natexlab{f}})}\BibitemShut {NoStop}%
\bibitem [{hin(2017{\natexlab{g}})}]{hina7}%
  \BibitemOpen
  \href
  {https://www.deccanchronicle.com/nation/current-affairs/050717/wb-section-144-imposed-in-basirhat-post-violent-clashes-over-facebook-post.html}
  {\bibfield  {journal} {\bibinfo  {journal} {Deccan Chronicle}\ } (\bibinfo
  {year} {2017}{\natexlab{g}})}\BibitemShut {NoStop}%
\bibitem [{hin(2017{\natexlab{h}})}]{hina8}%
  \BibitemOpen
  \href
  {https://www.hindustantimes.com/india-news/bengal-violence-over-fb-post-basirhat-peaceful-blockades-lifted-after-meeting-between-police-locals/story-c4xyOMzHMYZIWMxGMlvd9H.html}
  {\bibfield  {journal} {\bibinfo  {journal} {Hindustan Times}\ } (\bibinfo
  {year} {2017}{\natexlab{h}})}\BibitemShut {NoStop}%
\bibitem [{hin(2017{\natexlab{i}})}]{hina9}%
  \BibitemOpen
  \href
  {https://economictimes.indiatimes.com/news/politics-and-nation/social-media-posts-trigger-seven-communal-riots-in-a-month-in-west-bengal/articleshow/59496771.cms}
  {\bibfield  {journal} {\bibinfo  {journal} {The Asian Age}\ } (\bibinfo
  {year} {2017}{\natexlab{i}})}\BibitemShut {NoStop}%
\bibitem [{hin(2017{\natexlab{j}})}]{hina10}%
  \BibitemOpen
  \href
  {http://www.newindianexpress.com/nation/2017/jul/07/west-bengal-women-at-forefront-of-basirhat-violence-1625577.html}
  {\bibfield  {journal} {\bibinfo  {journal} {The New Indian Express}\ }
  (\bibinfo {year} {2017}{\natexlab{j}})}\BibitemShut {NoStop}%
\bibitem [{hin(2017{\natexlab{k}})}]{hina11}%
  \BibitemOpen
  \href
  {https://eisamay.indiatimes.com/west-bengal-news/hindu-muslim-family-proved-their-communal-harmony-at-basirhat/articleshow/59538641.cms?wesi=1}
  {\bibfield  {journal} {\bibinfo  {journal} {Eisamay (in Bengali)}\ }
  (\bibinfo {year} {2017}{\natexlab{k}})}\BibitemShut {NoStop}%
\bibitem [{hin(2017{\natexlab{l}})}]{hina12}%
  \BibitemOpen
  \href
  {https://www.dnaindia.com/india/report-no-fake-news-please-kolkata-police-arrest-rumour-monger-who-morphed-bhojpuri-film-to-spread-hate-2495917}
  {\bibfield  {journal} {\bibinfo  {journal} {Dna}\ } (\bibinfo {year}
  {2017}{\natexlab{l}})}\BibitemShut {NoStop}%
\bibitem [{hin(2017{\natexlab{m}})}]{hina13}%
  \BibitemOpen
  \href
  {https://www.deccanherald.com/content/621476/riot-hit-basirhat-tense-control.html}
  {\bibfield  {journal} {\bibinfo  {journal} {Deccan Herald}\ } (\bibinfo
  {year} {2017}{\natexlab{m}})}\BibitemShut {NoStop}%
\bibitem [{hin(2017{\natexlab{n}})}]{hina14}%
  \BibitemOpen
  \href {https://www.altnews.in/vicious-cycle-fake-images-basirhat-riots/}
  {\bibfield  {journal} {\bibinfo  {journal} {Altnews}\ } (\bibinfo {year}
  {2017}{\natexlab{n}})}\BibitemShut {NoStop}%
\bibitem [{Note1()}]{Note1}%
  \BibitemOpen
  \bibinfo {note} {In some village there were incidents of bike burning of
  Hindu youth. On the spread of this news to some other village, Muslim youths
  in that village were beaten up. In this way of local communication the riot
  began to spread. Name: Hindu Salesman; Address: Vill- Choto Jirat, Basirhat;
  Interview date, time: 20th October, 2019, 1:30pm at Itinda Ghat(near
  Icchamati River)}\BibitemShut {NoStop}%
\bibitem [{Note2()}]{Note2}%
  \BibitemOpen
  \bibinfo {note} {There were random rumours of attacks the madrasa or temples
  in the nearby villages. These rumours were propagating the riots. Name:
  Jallaluddin; Address: Bhabla halt, Basirhat; Occupation: Cycle garage owner;
  Interview date, time: 20th October, 2019, 3:00pm at Bhabla halt Station
  (Muslim area)}\BibitemShut {NoStop}%
\bibitem [{Note3()}]{Note3}%
  \BibitemOpen
  \bibinfo {note} {There were rumours about the attacks upon mandir-masjid but
  on his verification he found that there were no such incidents. Name: Tapas
  Ghosh; Address: Paikpara, Basirhat; Occupation: Shopkeeper; Interview date,
  time: 20th October, 2019, 4:00pm at Paikpara (Muslim area)}\BibitemShut
  {NoStop}%
\bibitem [{Note4()}]{Note4}%
  \BibitemOpen
  \bibinfo {note} {As he said a sick Muslim-youth along with his old mother
  were found sitting near the road. Out of riot fear, they were clueless about
  what to do and how to reach home. The Muslim-youth was discharged from
  hospital and was seen carrying saline bottles along with him. The Hindu
  salesman felt very sympathetic towards the Muslim-youth, he took the
  Muslim-youth to his house to provide him shelter. Due to this attitude the
  Hindu-neighbour of the Hindu-salesman at first got angry but later on they
  understood the situation. Name: Hindu Salesman; Address: Vill- Choto Jirat,
  Basirhat; Interview date, time: 20th October, 2019, 1:30pm at Itinda
  Ghat(near Icchamati River)}\BibitemShut {NoStop}%
\bibitem [{Note5()}]{Note5}%
  \BibitemOpen
  \bibinfo {note} {After the attack a Muslim doctor generously gave Rs. 5000/-
  to a poor Hindu shopkeeper whose shop was attacked. Name: Tapas Ghosh;
  Address: Paikpara, Basirhat; Occupation: Shopkeeper; Interview date, time:
  20th October, 2019, 4:00pm at Paikpara (Muslim area)}\BibitemShut {NoStop}%
\bibitem [{Note6()}]{Note6}%
  \BibitemOpen
  \bibinfo {note} {There exists a dependency of Hindu-Muslim in the area, it is
  seen that Muslim workers work under Hindu owner and vice-versa, and also seen
  that they work together as workers, that is the main reason of early
  convergence of the riot. Name: Sambhu Seth; Address: Trimohini, Basirhat;
  Occupation: Owner of Private Eye Hospital; Interview date, time: 20th
  October, 2019, 12:00pm}\BibitemShut {NoStop}%
\bibitem [{Note7()}]{Note7}%
  \BibitemOpen
  \bibinfo {note} {On 3rd and 4th day of July 2017 there was attack and loot in
  the Hindu shops of the area, along with a small pan and cigarette shop
  nearby. He claimed the loot was of 22lakhs during the incident. Out of sorrow
  he expressed that the small pan and cigarette shop received its reimbursement
  whereas others including him did not receive. In the Trimohini area the Hindu
  and Muslim shops are situated on opposite sides of the main road, due to
  which it is easy to identify the shops by their religion. In the evening at
  around 8pm of 3rd July there was an attack in the Rath Jatra at Trimohini
  which was followed by the loot at mid-night. Following this incident there
  were attacks and loots in the Muslim shops of the area on 5th July. Name:
  Sambhu Seth; Address: Trimohini, Basirhat; Occupation: Owner of Private Eye
  Hospital; Interview date, time: 20th October, 2019, 12:00pm}\BibitemShut
  {NoStop}%
\bibitem [{Note8()}]{Note8}%
  \BibitemOpen
  \bibinfo {note} {He said that the Hindu and Muslim residents of the area were
  united and wanted peace. He further accused towards the Bangladeshi Hindu
  refugees of the area. He also added that these Bangladeshi Hindu refugees
  have no permanent residence or land and mostly inhabits in the bustees of
  adjoining railway land. Along with these Bangladeshi Hindu refugees may be
  outsiders were involved in the riot. Name: Jallaluddin; Address: Bhabla halt,
  Basirhat; Occupation: Cycle garage owner; Interview date, time: 20th October,
  2019, 3:00pm}\BibitemShut {NoStop}%
\bibitem [{Note9()}]{Note9}%
  \BibitemOpen
  \bibinfo {note} {He said that there was no such Hindu-Muslim clash incident
  in Baduria not like Bashirhat. Mainly there was clash between police and the
  local people. Name: Muslim-Shopkeeper; Address: Baduria Chowmata (near
  Baduria police station); Interview date, time: 20th October, 2019,
  6:00pm}\BibitemShut {NoStop}%
\end{thebibliography}%
\end{document}